\documentclass[fleqn,usenatbib]{mnras}

\usepackage{newtxtext,newtxmath}

\usepackage[T1]{fontenc}

\DeclareRobustCommand{\VAN}[3]{#2}
\let\VANthebibliography\thebibliography
\def\thebibliography{\DeclareRobustCommand{\VAN}[3]{##3}\VANthebibliography}
\DeclareRobustCommand{\DEVAN}[3]{#2}

\def\thebibliography{\DeclareRobustCommand{\DEVAN}[3]{##3}\VANthebibliography}

\usepackage{graphicx}	
\usepackage{amsmath}	
\usepackage{xcolor}

\newcommand{\afe}{$[\alpha/\rm{Fe}]$}
\newcommand{\feh}{$[\rm{Fe}/\rm{H}]$}
\newcommand{\oxh}{$[\rm{O}/\rm{H}]$}
\newcommand{\msun}{M$_\odot$}

\newcommand{\rsun}{R$_\odot$}
\newcommand{\teff}{T$_{\mathrm{eff}}$}
\newcommand{\bpass}{{\sc{bpass}}}

\title[$\alpha$-enhanced BPASS single star models]{BPASS stellar evolution models incorporating $\alpha$-enhanced composition -- \\ I. Single star models from $0.1$ to $316$\,\msun}

\author[Byrne et al.]{
Conor M. Byrne,$^{1}$\thanks{E-mail: conor.byrne@warwick.ac.uk (CMB)}
Jan J. Eldridge$^{2,1}$
and Elizabeth R. Stanway$^{1}$
\\
$^{1}$Department of Physics, University of Warwick, Gibbet Hill Road, Coventry CV4 7AL, UK\\
$^{2}$Department of Physics, University of Auckland, Private Bag 92019, Auckland, New Zealand}

\date{Accepted 28 January 2025. Received 27 January 2025; in original form 01 August 2024}

\pubyear{2024}

\begin{document}
\label{firstpage}
\pagerange{\pageref{firstpage}--\pageref{lastpage}}
\maketitle

\begin{abstract}
Stellar evolution modelling is fundamental to many areas of astrophysics including stellar populations in both nearby and distant galaxies. It is heavily influenced by chemical composition. Observations of distant galaxies and nucleosynthesis calculations show that $\alpha$-process elements are enriched faster than iron group elements. We present a dense grid of single-star models calculated using the \bpass\ stellar evolution code and covering masses ($0.1\le\mathrm{M}/$\msun$\le316$), metallicity mass fractions ($10^{-5} \le Z \le 0.04$) and $\alpha$-to-iron abundance ratios ($-0.2\le$\afe$\le+0.6$). By comparing Solar-scaled models to ones enriched in $\alpha$-process elements, we find that stellar radii, surface temperatures, Main Sequence lifetimes, supernova progenitor properties and supernova rates are all sensitive to changes in \afe. Lifetimes of low-mass stars differ by up to 0.4 dex, while surface temperatures of massive stars at the end of the Main Sequence also differ by around 0.4 dex.  Inferred supernova rates when \feh\ is unknown can be highly uncertain.
Models with different \afe\ but comparable iron abundances show smaller variations, indicating that while iron primarily defines the course of evolution; $\alpha$-enhancement nonetheless has an impact of up to 0.1 dex on stellar properties. Such changes are small for individual stars, but have a large cumulative effect when considering an entire stellar population as demonstrated by isochrone fitting to nearby clusters. Changes in radii and lifetimes have further consequences for a stellar population including binary stars, as they influence the timing, nature and occurrence rate of mass transfer events. 


\end{abstract}




\section{Introduction}

The evolution of a star can be characterised to the first order by its mass and chemical composition, which dictate the amount of fuel available for nuclear fusion and the rate at which these reactions take place. A more accurate picture can be drawn by considering the micro-physical and macro-physical processes in more detail: processes such as mass loss, convection and radiative transfer. Many uncertainties remain in our understanding of stellar evolution \citep[e.g.][]{Stancliffe16,Agrawal20}. In particular, the evolution of massive stars remains quite uncertain, with large differences found between different stellar evolution models \citep[][]{Agrawal22,Eldridge22Review}.

The mass $M_*$ and metallicity mass fraction $Z$ are the most important inputs to a computational stellar evolution model as they play the largest role in determining how a star will change over time. In most widely used stellar evolution codes, $Z$ is defined relative to the elemental composition of the Sun. Numerous measurements have been made of the Solar composition, each slightly differing from one another \citep[e.g.][]{Grevesse93,Grevesse98,Asplund05,Asplund09,Asplund21}. Regardless of the chosen definition of Solar composition, the most common approach for generating a stellar evolution model with a sub-Solar value of $Z$ is to take the Solar composition and scale the elemental abundances uniformly to reach the desired metal content.

One aspect which is rapidly requiring attention is the relative abundances of different chemical elements in stars, since in stellar populations in the early Universe, low-mass stars have not had sufficient time to reach the end of their lives and enrich their environment with iron and related elements. Conversely, the deaths of shorter-lived massive stars in core-collapse supernovae (CCSNe) enrich their environment in oxygen and other $\alpha$-process elements \citep[e.g.][]{Kobayashi20}. This behaviour, commonly referred to as $\alpha$-enhancement is widely observed in old stellar populations in the nearby Universe \citep[i.e. populations of stars which formed when the Universe was young, before significant iron enrichment can take place, e.g.][]{Larsen18} as well as in stellar populations at moderate and high redshift, where the Universe is simply too young for the low-mass stars to have fully evolved \citep[e.g.][]{2021MNRAS.505..903C}.

With deep surveys, JWST is discovering galaxies at very high redshifts \citep[e.g.][]{JWSTAdams23,JWSTAtek23,JWSTCastellano22,JWSTDonnan23,JWSTFinkelstein22,JWSTFujimoto23,JWSTMcLeod24,JWSTNaidu22,JWSTYan23}. The light from these galaxies was emitted just a few hundred million years after the Big Bang, and thus their stellar populations are dominated by hot, young massive stars. Consequently, the standard approach of employing stellar models with Solar-scaled composition are not necessarily suitable in this regime. This raises a number of challenges for existing theoretical models, where the impact of non-Solar-scaled compositions have not been well explored, particularly in the context of massive stars in the early Universe.

Some studies have looked at the impact of non-uniform composition scaling on stellar evolution models. \citet{Grasha21} calculated stellar evolution models for stars between 10 and 300\,\msun with metallicity following the Galactic Concordance chemical abundance scheme. From these models, they conclude that while the chemical abundance scheme has limited impact on stars at or close to Solar metallicity, significant variation in evolutionary behaviour can be seen at low metallicity, particularly at high mass. They also find that stellar rotation in massive stars impact their ionising photon output to a similar degree as abundance changes at low metallicity.

\citet{Farrell22} computed detailed stellar evolution models exploring the effects that changes to the interior composition and stellar interior physics have on massive stars. They find that low metallicity massive stars tend to have a higher \teff\ as the lower internal CNO abundances affect nuclear energy generation and reduce the opacity of the stellar envelope. 

Overall, the findings of these works illustrate that small changes to the assumed interior structure and composition of a star can have a large affect on their surface properties and evolution. With observational evidence clearly indicating $\alpha$-enhancement in low-metallicity stellar populations and theoretical models showing that small changes in composition can manifest large changes in stellar surface properties, it is readily apparent that large grids of stellar evolution models, such as those used for stellar population and spectral synthesis, need to account for these variations in elemental composition also.

Since enrichment is a consequence of the differing timescales between CCSNe and other, slower processes, the impact of $\alpha$-enhancement on SN rate is also of interest here. Previous work has considered the impact of bulk metallicity variations on SN rates and the limiting masses of stars undergoing core-collapse \citep[e.g.][]{Eldridge08,2008ApJ...673..999P,2009A&A...503..137B,2023ApJ...955L..29P}. However the lack of detailed massive star models with varying composition has prevented further analysis.

In this work we present a grid of single-star stellar evolution models calculated within the \bpass\ (Binary Population And Spectral Synthesis) framework, covering a broad range of compositions and stellar masses. We examine the impact that $\alpha$-enhancement has on stellar evolution and observable properties of stars, with an emphasis on massive stars and applications to the high-redshift Universe. We also consider the impact of composition on the minimum mass of stars to undergo core collapse, and the consequences for the inferred SN rate. Based on the single-star results, we discuss the potential impacts for binary stellar populations, spectral synthesis and consequences for the distant Universe. 

\section{BPASS stellar evolution modelling}
\subsection{BPASS evolution code}
\label{sec:stelmod}

The \bpass\ stellar evolution code is a development of the Cambridge STARS code and was last described in detail in \citet{2017PASA...34...58E}, with the release of \bpass\ v2.1. Improvements and modifications were made in subsequent years leading to the release of v2.2.1 \citep{2018MNRAS.479...75S}. Changes to the stellar spectral libraries were made in order to include $\alpha$-enhanced compositions in the release of v2.3 \citep{Byrne22}, but the underlying stellar evolution models have thus far remained Solar-scaled.

The stellar evolution models presented in this work will only consider isolated, single-star evolution, and thus do not comprise a full release of \bpass. However we identify the set of models presented here as \bpass\ v2.4.0.
Detailed binary stellar evolution calculations are ongoing and will form the basis of future publications. Nonetheless, here we restate the key stellar physics choices that have been made to generate this set of models and outline the modifications that have been made since \citet{2018MNRAS.479...75S}. 

A key change has been the added capability to scale elemental abundances in a non-uniform manner, in order to properly compute stellar evolution models with $\alpha$-enhanced compositions. To simply update the initial compositions of the stellar models is relatively straightforward. However, we must also allow for the changes the $\alpha$-enhanced composition causes to the opacity of the stellar envelope. To do this we use the opacity tables as originally computed in \citet{Eldridge04a}. These tables are built from the OPAL \citep[][]{Iglesias96} opacity tables, in the Type 2 format that accounts for changes in H, He, C and O. Since oxygen is the $\alpha$-process element that contributes most strongly to the stellar opacity, this flexibility enables appropriate, tabulated opacity data to be utilised for $\alpha$-enhanced compositions. 

While implementing this change, the Solar $\alpha$-to-iron ratio was redefined such that the base \citet{Grevesse93} elemental mixture -- which is more oxygen-rich than most other and more recent definitions of the Solar composition -- is now defined to have an \afe$=+0.2$. As a consequence the fiducial \bpass\ abundances used in v2.3 and earlier are equivalent in composition to the \afe$=+0.2$ models that we present here, and the `Solar' composition we use is an $\alpha$-depleted version of the \citet{Grevesse93} abundances.

In the model set presented here we calculate 15 initial bulk metallicity mass fractions at $Z=10^{-5}$, $10^{-4}$, $3\times10^{4}$, 0.001, 0,002, 0.003, 0.004, 0.006, 0.008, 0.010, 0.014, 0.017, 0.020, 0.030 and 0.040. For each metallicity the hydrogen mass fraction is computed as $X=0.75-2.5Z$ and the helium mass fraction as the remainder. Within these bulk metallicities, we distribute the metals between C, N, O, Ne, Mg and Fe as shown in Table~\ref{tab:alpha_abundances} for each of the $\alpha$-enhancements. Thus to calculate the mass fraction of iron for a certain metallicity and $\alpha$ composition we simply multiply the bulk metallicity by the value tabulated.

This leaves us with 74 different compositions\footnote{An opacity table with a high enough iron abundance to satisfy $Z=0.040$ and \afe=-0.2 does not exist, so we are unable to compute this model grid.} defined by $\alpha$ composition and bulk metallicity. We interpolate between the opacity tables calculated by \citet{Eldridge04a} to obtain a table with the same iron abundance as that of the required model composition in each case.

One caveat that arises from this approach is that the original opacity tables assume a minimum fraction of the metals that can be attributed to carbon and oxygen, corresponding to the abundance ratio used at \afe$=+0.2$. The tables are not recalculated below this value. In the grids with \afe$<0.2$, there is a risk that opacities are slightly overestimated. However in most circumstances where the opacity is dominated by metal opacity (hydrogen-rich and helium-rich stellar envelopes at high temperatures), iron-group elements are the largest contributor. In addition, this happens in all stellar models when CNO processing redistributes carbon and oxygen to nitrogen and is thus a common assumption. The spacing of the opacity grids in excess carbon and oxygen \citep[see][for details]{1993ApJ...412..752I} is greater than the differences in abundances from varying \afe. Thus, until core nucleosynthesis boosts the carbon and oxygen abundances significantly above the initial values, their contribution to the total opacity is comparatively unimportant. At that point the composition is sufficiently rich in carbon and oxygen to use the CO-rich tables in the Type 2 format opacity data. Thus, at the levels of variation in carbon and oxygen abundances we accommodate within our grids, the potential error introduced is small.

In addition to the compositions and opacities, we have made other refinements to the stellar evolution physics. We have updated two of the nuclear reaction rates. In light of the work of \citet{2022ApJ...931..157A}, we have updated the $^{12}$C$+\alpha\rightarrow ^{16}$O to that given by \citet{2002ApJ...567..643K}. Using this rate within the \bpass\ evolution code allows us to reproduce observed Wolf-Rayet star surface compositions with greater fidelity \citep{2022ApJ...931..157A}. In addition, we update the $^{12}$C$+^{12}$C reactions to the \textit{HinRes} rates from \citet{2022A&A...660A..47M}. This adjustment accounts for the impact of a resonance in this reaction on the rate.

Finally, we have updated the mass-loss rate prescription to include a relevant mass loss rate for asymptotic giant branch (AGB) stars within the evolution code. Since \bpass\ v2.2 we have implemented this additional mass loss as a post-processing step as outlined in \citet{2018MNRAS.479...75S}. It is now fully integrated to the main stellar evolution routines. The scheme we use is similar to that in \citet{1995A&A...297..727B,1995A&A...299..755B}. First, we modify the mass-loss rates for red giant branch stars. For stars with an initial mass less than 8~M$_{\odot}$, a helium core that is smaller than 2.5~M$_{\odot}$ and a effective temperature less than 5200~K we use a  \citet[][]{Reimers75}-like mass-loss rate with,
\begin{equation}
\dot{M}_{\rm RGB} = 4\times 10^{-12} \frac{L}{L_{\odot}} \frac{R}{R_{\odot}} \frac{M_{\odot}}{M} M_{\odot}\,{\rm yr^{-1}}
\end{equation}
The period of pulsations for the giants are then computed as
\begin{equation}
\log(P/{\rm days}) = -1.92 -0.73\log(M/M_{\odot})+1.83\log(R/R_{\odot}).
\end{equation}
When the period is greater than 100~days and the difference in mass coordinate of the hydrogen and helium burning shells is less than 0.1~M$_{\odot}$ we switch to an AGB mass-loss rate prescription given by
\begin{equation}
\dot{M}_{\rm AGB} = 3.864\times10^{-21} \left(\frac{L}{L_{\odot}}\right)^{3.7} \frac{R}{R_{\odot}} \left(\frac{M_{\odot}}{M}\right)^{-3.1} M_{\odot}\,{\rm yr^{-1}}.
\end{equation}
When we are close to removing the hydrogen envelope, with less than 0.01~M$_{\odot}$ of hydrogen envelope remaining, we slowly reduce the mass-loss rates to zero as the stellar model becomes a white dwarf.
These changes to the mass-loss prescription mean that many of our models now naturally become white dwarfs towards the end of their evolution, although some models still encounter numerical difficulties at the tip of the AGB due to the rapidly evolving (in time and mass co-ordinate) nature of the thin burning shells and the thermal pulses, which is a challenging phase for 1-dimensional stellar evolution codes to compute \citep[e.g.][]{Rees24}.

In other respects, the physics of the \bpass\ stellar models remains unchanged from previous descriptions.

\begin{table}
 \caption{Mass fraction of each element as a fraction of the total bulk metallicity, for each $\alpha$ composition. Only these elements are followed as a composition variable within the evolution code.}
 \label{tab:alpha_abundances}
 \begin{tabular}{rccccc}
  \hline
  Element &  &   &  &   &  \\
  Mass     & \afe & \afe &\afe & \afe &\afe \\
  Fraction & =+0.6 & =+0.4 & =+0.2 & =+0.0 & =-0.2\\ 
  \hline
$X_\mathrm{C}/Z$ & 0.0877 & 0.1258 & 0.1733 & 0.2275 & 0.2836  \\
$X_\mathrm{N}/Z$  & 0.0269 & 0.0386 & 0.0532 & 0.0698 & 0.0870 \\
$X_\mathrm{O}/Z$  & 0.6128 & 0.5547 & 0.4823 & 0.3996 & 0.3142 \\
$X_\mathrm{Ne}/Z$ & 0.1254 & 0.1135 & 0.0987 & 0.0817 & 0.0643 \\
$X_\mathrm{Mg}/Z$ & 0.0477 & 0.0432 & 0.0376 & 0.0311 & 0.0245 \\
$X_\mathrm{Si}/Z$ & 0.0205 & 0.0294 & 0.0405 & 0.0532 & 0.0663  \\
$X_\mathrm{Fe}/Z$ & 0.0363 & 0.0521 & 0.0718 & 0.0943 & 0.1175  \\
  \hline
 \end{tabular}
\end{table}

\subsection{Single star grid}

Stellar evolution models were calculated for single stars from $\log(\mathrm{M}/$\msun$)=-1$ (0.1 \msun) to $\log(\mathrm{M}/$\msun$)=2.5$ (316 \msun) with logarithmic spacing $\Delta\log(\mathrm{M})=0.01$, yielding 351 mass points. Models were calculated for 15 different values of bulk metallicity mass fraction, $Z$ and 5 values of \afe, as outlined above. Each composition is referred to by its unique combination of $Z$ and \afe\ (e.g. {\texttt{z020\_a+0.0}} corresponds to the fiducial \bpass\ composition of $Z=0.020$ and \afe=0.0). This arrangement of elemental mixtures has the natural consequence that models with the same $Z$ which are $\alpha$-enhanced will be iron-poor relative to the Solar-scaled abundance pattern and vice versa.

The primary motivation for $\alpha$-enhanced stellar models in this work is to better understand stellar populations in the early Universe, where the contribution of massive stars (M$\gtrsim8$\msun) dominate the emitted light. Nonetheless, we will comment on notable trends and behaviours that are seen at all masses. We will consider the impact of three different ways of changing chemical composition:
\begin{enumerate}
    \item Changes to the bulk metallicity mass fraction $Z$, keeping the elemental abundance ratios fixed (i.e. the canonical method for scaling metal abundances)
    \item Changes to \afe\ while keeping $Z$ fixed. As outlined above, this means for a given $Z$, the abundance of Fe decreases as $\alpha$ elements increase
    \item Choosing combinations of $Z$ and \afe\ which give an equivalent mass fraction of iron. Since iron is known to be a key contributor to stellar opacity, changes in iron are expected to be the main variable determining the evolution of a star at fixed initial mass.
\end{enumerate}

We will consider scenarios (i) and (ii) in Sections~\ref{sec:resultmain} and~\ref{sec:PMS}, while scenario (iii) will be briefly examined in Section~\ref{sec:fixFe}. A bulk metallicity of $Z=0.020$ and \afe=0.0 represent the fiducial \bpass\ composition, although it should be noted that given the reassignment of abundance patterns outlined in Section~\ref{sec:stelmod} above, this corresponds to a slightly $\alpha$-depleted version of the \citet{Grevesse93} abundances used in earlier versions of \bpass.

\section{Main Sequence evolution}\label{sec:resultmain}
\subsection{Evolution tracks at selected masses}
\label{sec:deltaZ}

\begin{figure*}
    \centering
    \includegraphics[width=0.97\textwidth]{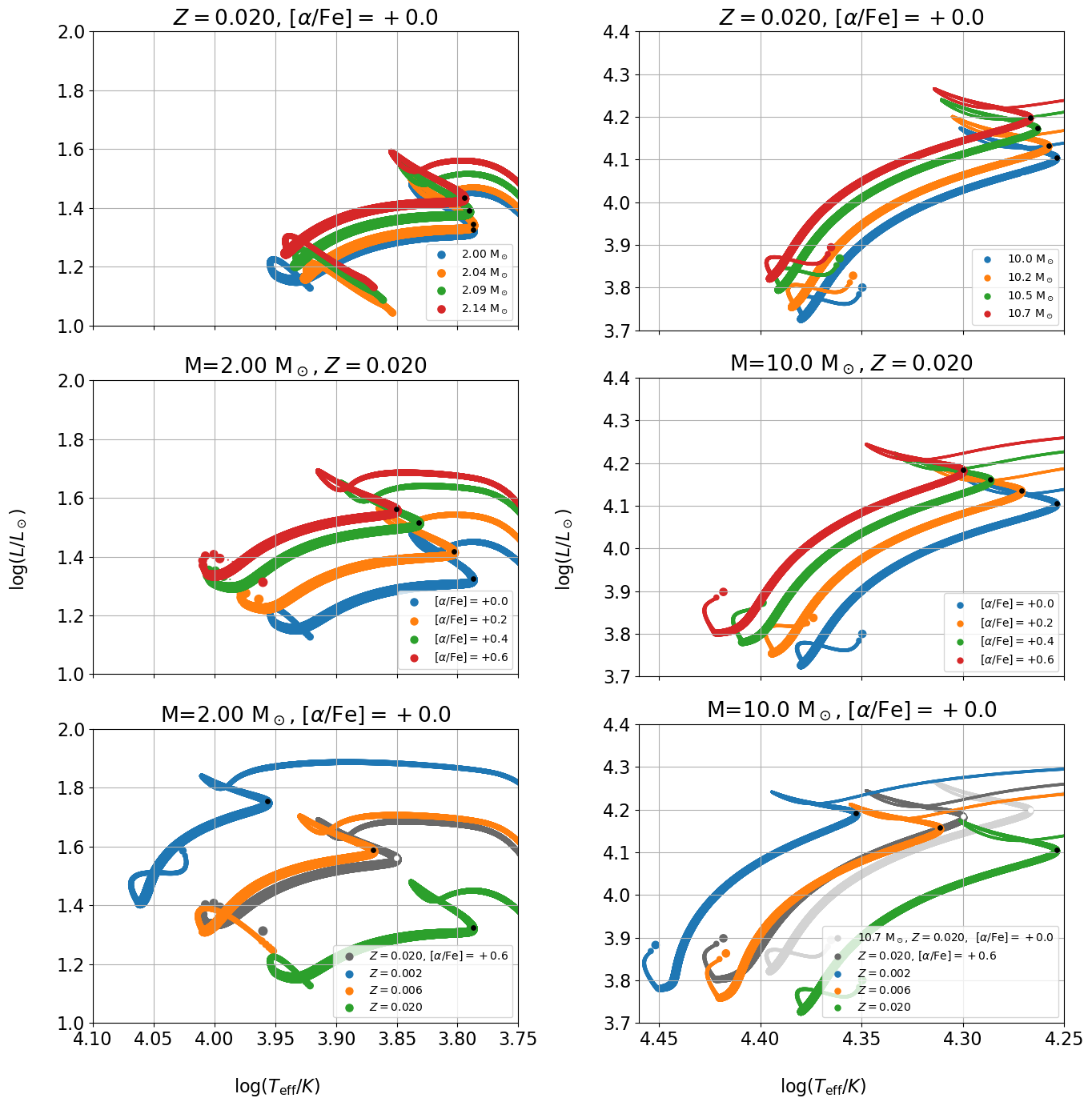}
    \caption{Comparative HRDs for stellar evolution models of 2.0 \msun\ (left panels) and 10.0 \msun\ (right panels). The thickness of the lines corresponds to the duration spent at each point, indicating periods of slow (thick line) and rapid (thin line) evolution. The coloured lines show how changes in specific parameters influence the evolution track, the black (white) points indicate the point at which the central hydrogen mass fraction drops below 10 per cent in the coloured (grey) tracks. {\emph{Upper panels}}: Changes in initial stellar mass, with a fixed composition of $Z=0.020$ and \afe$=0.0$. {\emph{Middle panels}}: Mass fixed at the reference value, total metallicity fixed at $Z=0.020$, but with variations in \afe. {\emph{Lower panels}}: The coloured lines show models of the same reference mass, a fixed value of \afe$=0.0$, but different total metallicity. The light grey and dark grey lines highlight the tracks of a model with increased mass and a model with increased \afe\ respectively, taken from the panels above, with properties as indicated by the legend. Each column of panels uses the same range on the axes for direct comparison.}
    \label{fig:hr1020}
\end{figure*}

\begin{figure*}
    \centering
    \includegraphics[width=0.97\textwidth]{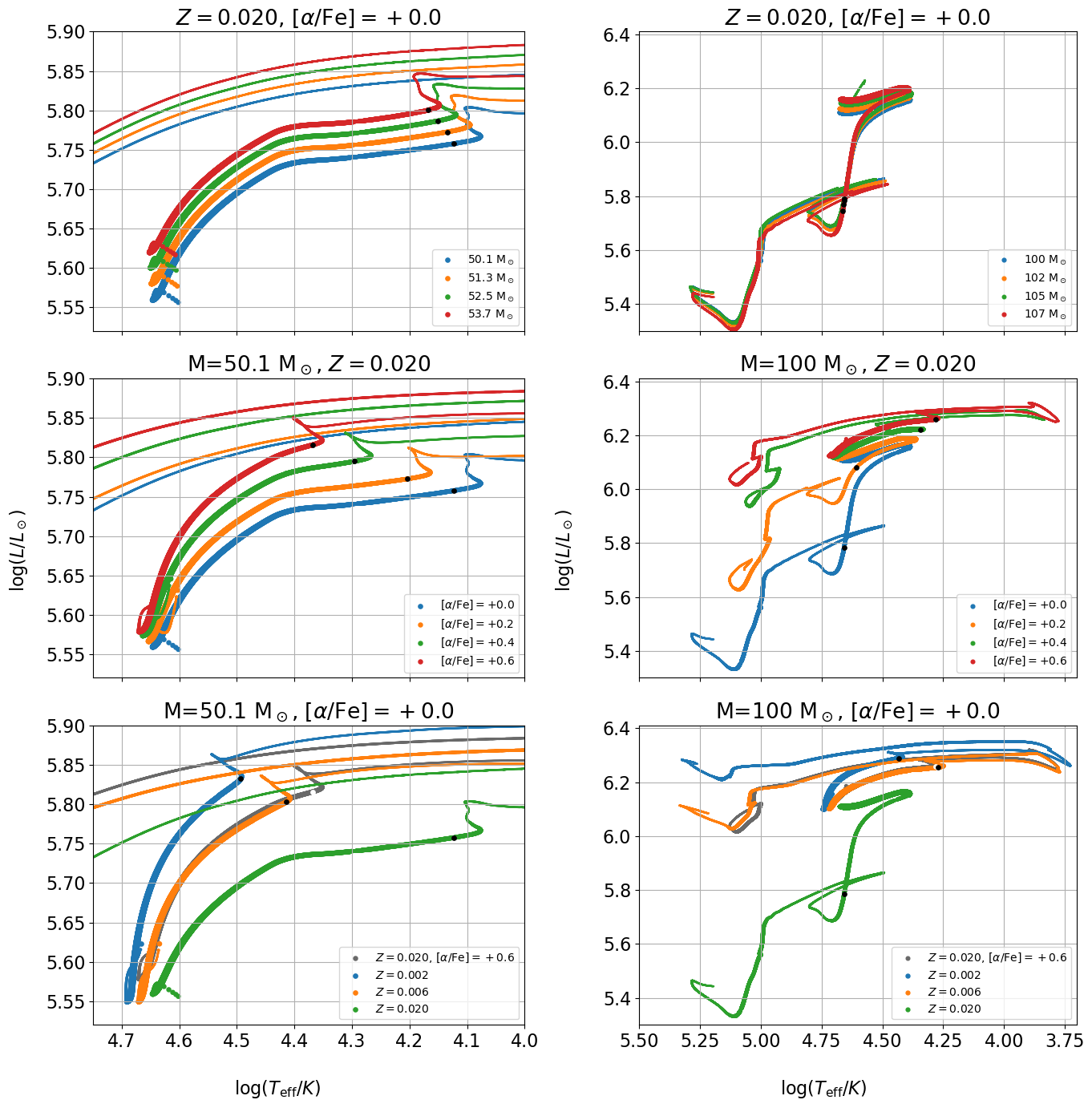}

    \caption{As Fig~\protect\ref{fig:hr1020}, but for reference models with an initial mass of 50.1 \msun\ (left panels) and 100 \msun\ (right panels).}
    \label{fig:hr50100}
\end{figure*}

First we shall consider the evolution of models at specific masses and their location in the Hertzsprung-Russell diagram (HRD). Evolution tracks for reference models with initial masses of 10.0, 20.0 are shown in Fig~\ref{fig:hr1020} and models with initial masses of 50.1 and 100 \msun\ are shown in Fig~\ref{fig:hr50100}. Each panel shows an enlarged portion of the HRD centred on the Main Sequence evolution of the models. The upper panels show models with a small variation in initial mass, but a fixed composition ($Z=0.020$,\afe$=+0.0$), the middle panels compare models of varying values of \afe, but fixed initial mass and $Z=0.020$, while the lower panels present tracks associated with models where $Z$ has been varied but the initial mass has been held fixed and \afe$=+0.0$. One higher mass model and one $\alpha$-enhanced model from the panels above are plotted in light grey and dark grey respectively for comparison. Given that the changes in mass and \afe\ produce smooth variations in the corresponding panels (with the exception of the 100 \msun\ model), these provide a clear indication of the direction of motion in the luminosity-effective temperature plane arising from increases in mass and $\alpha$-enhancement respectively. The location of the terminal-age Main Sequence (TAMS) is indicated by the black and white dots on each of the coloured and grey tracks respectively. In this work we have defined the location of the TAMS as the point at which the core hydrogen mass fraction drops below 2.5 per cent. This generally corresponds to the `hook' in the HRD at the end of the Main Sequence evolution. While this does not always strictly correspond to the point at which a star evolves off the Main Sequence, it provides a fixed point at the end of the Main Sequence at which we can compare models of differing composition.

Looking at the location of the TAMS in each case, the effect that each change has on the stellar model and its Main Sequence evolution can be noted. Changes in mass (upper panels) move the tracks predominantly in the vertical direction, with higher masses moving to higher luminosities and little or no change in effective temperature. This follows the direction in which the Main Sequence travels at these masses, as one would expect. For example, we see an increase in the TAMS luminosity of about 0.042 dex between models of 50.1 and 53.7 \msun, an increase of 0.03 dex in $\log(M)$. The change in temperature at this point is similar at about 0.044 dex.

Changes to the total metallicity $Z$ (lower panels) without considering $\alpha$-enhancement lead to movement predominantly in the horizontal direction, with lower $Z$ leading to higher temperatures with only minor increases in luminosity. This is also expected behaviour, since a lower overall metallicity reduces the opacity of the stellar envelope, leading to higher surface temperatures and smaller radii. 
 
At the higher masses, the metallicity dependence of mass-loss rates becomes apparent. With lower metallicity, the lower mass-loss rates mean the star retains more mass for longer, thus staying bluer for more of its life, hence the much higher temperatures in the low $Z$ model.

\begin{figure}
    \centering
    \includegraphics[width=0.47\textwidth]{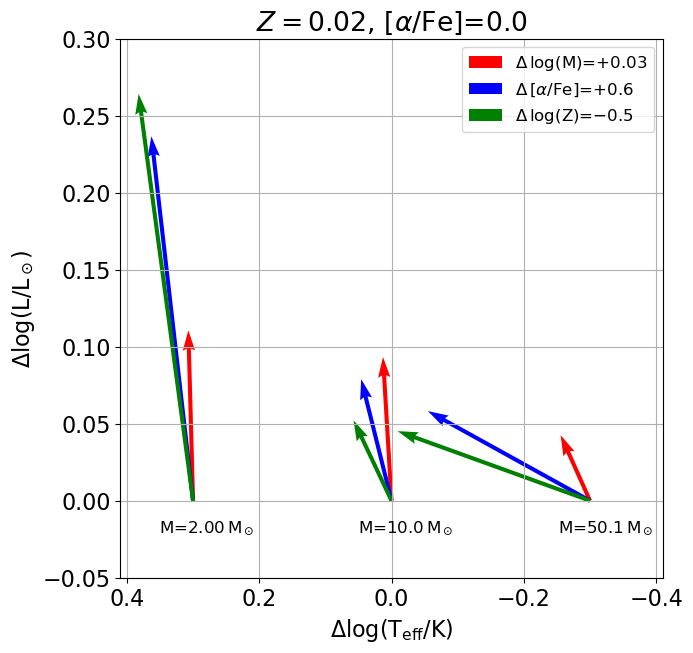}
    \caption{Vectors illustrating the movement in locus of TAMS on the log(L)--log(T$_{\mathrm{eff}}$) plane from changes in mass, \afe\ and $Z$ for reference models at 2.0\msun, 10.0\msun\ and 50.1\msun.}
    \label{fig:vec50}
\end{figure}

Changing \afe\ (centre panels) produces changes which lie somewhere between the two previous cases, with small shifts in both effective temperature and luminosity. Fig~\ref{fig:vec50} shows graphically the direction of motion of the TAMS luminosity and effective temperature for each of the changes: a small increase ($+0.03$ dex in log space) in mass, a change in metallicity from 0.020 to 0.006 ($-0.5$ dex) and a change in \afe\ from 0.0 to +0.6 starting from a model with a given mass and a Solar composition. The precise amount by which the composition and mass are changed is somewhat arbitrary, but chosen such that the length of the vectors are comparable, and in a range where the shift in TAMS locus is smooth, as informed by Figs~\ref{fig:hr1020} and~\ref{fig:hr50100}. This provides a visual indication of how a change in \afe\ leads to a shift in the effective temperature--luminosity plane which is intermediate to the shifts due to changes in mass and $Z$.

At 100 \msun, the response of the star to changes in mass, bulk metallicity and \afe\ are considerably different. This is due to the fact that the evolution of these very massive stars is dominated by mass-loss rates, which in turn are dependent on metallicity. Our \bpass\ stellar evolution models use the \citet{Vink01} mass-loss rates for hot Main Sequence stars (T$_{\mathrm{eff}}>25\,000$\,K), switching to \citet{Nugis00} rates for stripped, core-helium burning and Wolf-Rayet stars. Small changes in mass at fixed composition lead to very similar evolutionary tracks (upper panel), since the overall mass-loss history of each star is comparable. Reducing the iron content, either through lowering the bulk metallicity or through an increase in \afe\ at fixed $Z$, reduces the mass-loss rates of these stars. Thus, they retain more mass for longer, remaining more luminous over the course of their evolution. 

It is interesting to contrast the evolution of a model with Solar-scaled composition with a model with an $\alpha$-enhanced abundance profile, but a higher total $Z$. If chosen correctly, a subset of models can be chosen which have comparable iron abundances. Since iron is a key contributor to opacity in the stellar envelope, such a comparison will provide insight into the effect of increasing the $\alpha$ elements alone. 

This can be done by comparing the orange ($Z=0.006$, \afe$=+0.6$) and dark grey ($Z=0.020$, \afe$=+0.0$) tracks in the lower panels of Figs~\ref{fig:hr1020} and~\ref{fig:hr50100}. This comparison reveals that the Main Sequence evolution of models with equivalent iron content ($Z=0.006$, \afe=0 and $Z=0.020$, \afe$=+0.6$) follow very similar tracks, although the $\alpha$-enhanced models extend further beyond the TAMS turn-off seen in the Solar-scaled models. A further comparison of models with matching iron abundances is carried out in Section~\ref{sec:fixFe}.

\subsection{TAMS properties as a function of composition and mass}
\label{sec:TAMS}
\begin{figure*}
    \centering
    \includegraphics[width=0.98\textwidth]{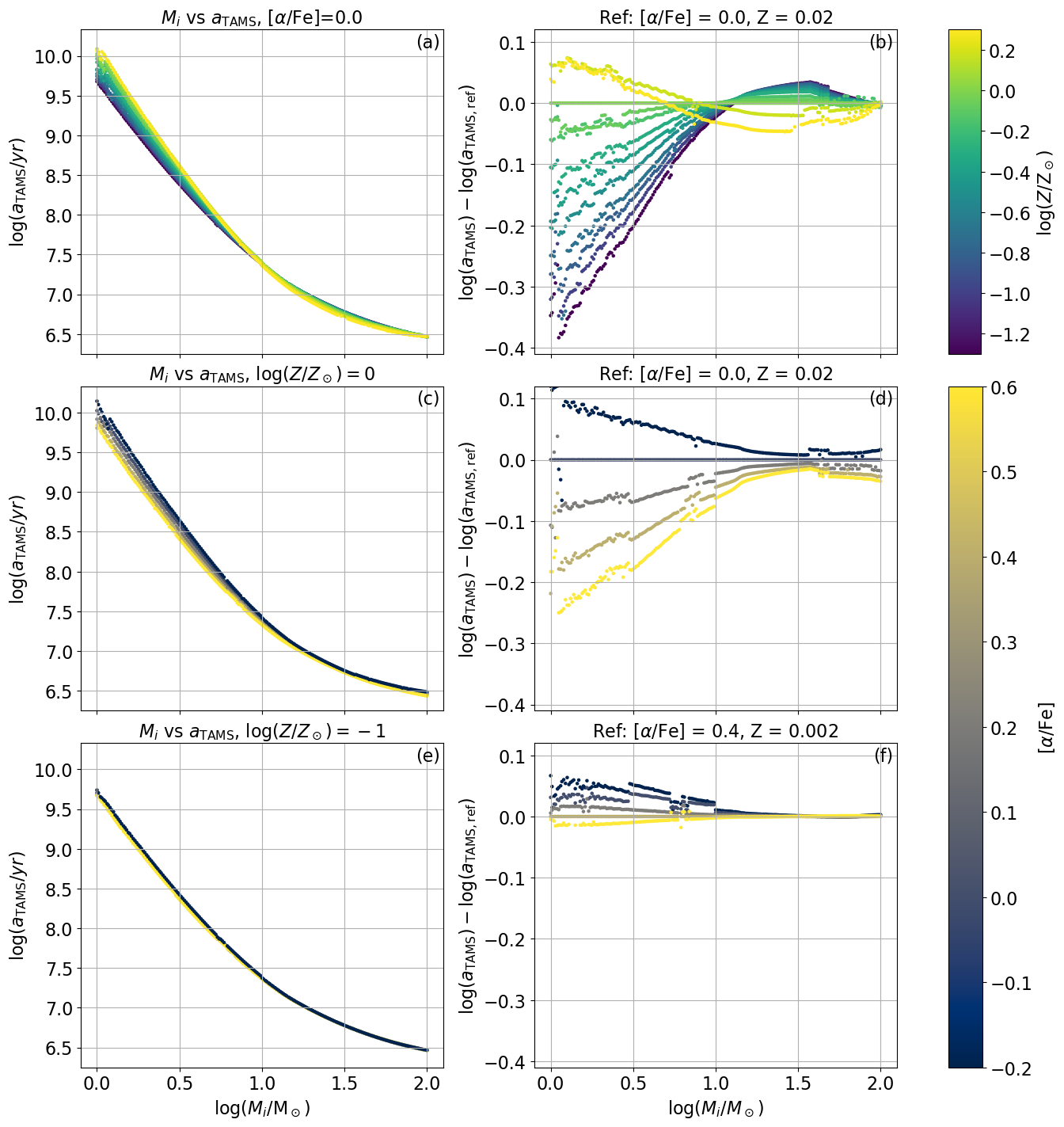}
    \caption{TAMS ages as a function of initial stellar mass. The left-hand panels show the true values, while the right-hand panels show the difference in log(age) relative to a reference composition, indicated by the relevant panel titles. Panels (a) and (b) show variations in metallicity mass fraction $Z$ for a Solar-scaled composition. Panels (c) and (d) show a Solar $Z$, with varying \afe. Panels (e) and (f) show ten per cent Solar models with varying \afe. The composition in each pair of panels is colour coded according to the colour scales on the right-hand side of the figure.}
    \label{fig:ATAMS}
\end{figure*}

\begin{figure*}
    \centering
    \includegraphics[width=0.98\textwidth]{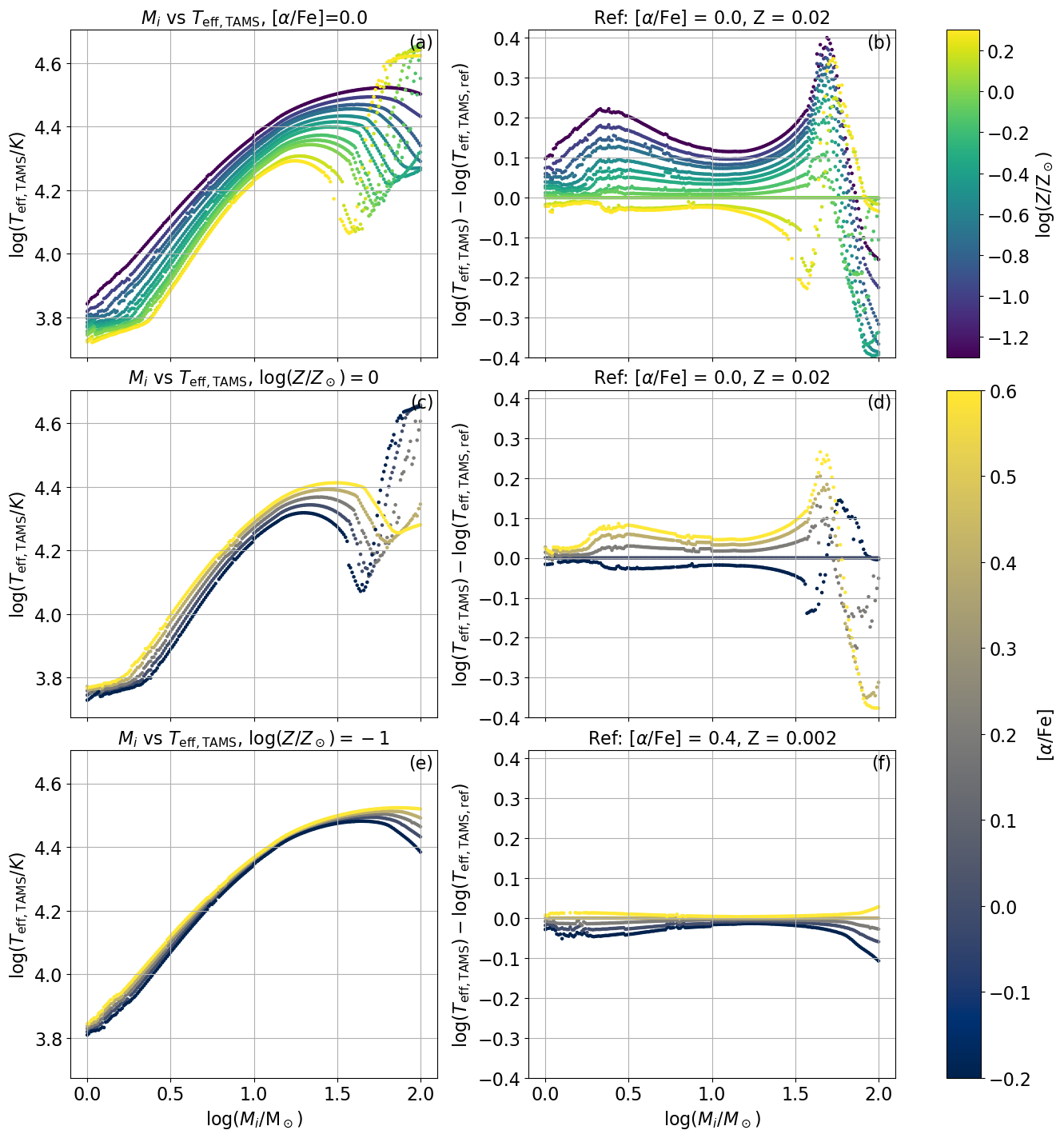}
    \caption{As Fig~\protect\ref{fig:ATAMS}, but for TAMS effective temperatures, rather than age.}
    \label{fig:TTAMS}
\end{figure*}

While the previous section has provided snapshots at individual masses and give an indication of the impact that composition changes have on HR evolution tracks, comparing properties across the full range of masses and compositions enables identification of properties and mass regimes which are particularly sensitive to composition changes. Additionally, the comparative HR diagrams presented above provide no clear indication of the actual age of the models. The TAMS point indicated on the tracks is fixed by the physical conditions in the core, but this does not correspond directly to a stellar age. Examining the stellar properties at the end of the Main Sequence provide an insight into the cumulative effect the different abundances have over the course of the star's core-hydrogen-burning life, the longest-lived phase of evolution.

Fig~\ref{fig:ATAMS} shows how TAMS age changes with initial stellar mass, with colours indicating the composition. The left-hand panels (a, c, and e) show the absolute values of log(age) as a function of mass while the right-hand panels (b, d, and f) illustrate the difference in log(age) from a stated reference composition. The top panels show changes in bulk metallicity (Z), the centre panels highlight changes in \afe\ at Solar Z (0.020), and the lower panels illustrate changes in \afe\ at 0.1 Z$_\odot$ (0.002). We limit these plots to the mass range of $1\leq M/$\msun$\leq 100$. The very low-mass models do not reach the TAMS before a Hubble time, while the very massive stars (VMS, above 100\msun) are subject to uncertainties in their mass-loss rates and their response to composition changes are less smooth than typical massive stars (as seen in Fig~\ref{fig:hr50100}, for example). 

As expected, changing bulk metallicity (upper panels) has a noticeable effect on Main Sequence lifetimes. For a Solar-scaled composition, this is particularly pronounced in the low-mass stars, where metal-poor stars display shorter lifetimes. This is a consequence of their lower opacity, which leads to more compact stars, higher pressures and higher nuclear reaction rates. Additionally, by losing more mass, the metal-rich stars can prolong their life relative to the metal-poor ones. The metal-poor stars remain more massive and more luminous for a greater fraction of their lifetime, consuming their fuel at a higher rate. Above $\sim$10 \msun, the situation is reversed, and the metal-poor stars show marginally longer lifetimes, due to the much steeper mass-to-age relation. These more massive, metal-poor/$\alpha$-enhanced stars remain hotter and more luminous throughout their lives, so much so that it counteracts the life-lengthening effects of mass-loss and compactness which give their low-mass counterparts greater ages than the Solar composition models.

A similar, less pronounced, trend can be seen for models with an $\alpha$-enhanced composition, with a broader mass range in which the metal-poor stars have a longer lifetime (M/\msun$\gtrsim 6$). Considering a fixed $Z$ but varying the elemental mixture, the changes in age are comparable in magnitude for $Z=0.020$, with a clear trend of shorter lifetimes for $\alpha$-enhanced compositions. The impact of $\alpha$-enhancement on lifetime is minimal at $Z=0.002$, with a maximum difference of 0.1 dex in age seen at 1\msun, and almost no difference for M/\msun>10.

As the stellar lifetimes span over 4 orders of magnitude, these relative differences may be more easily quantified if one plots the difference in $\log(\mathrm{age/yr})$ relative to a reference composition as shown in the right-hand panels.

Below 10 \msun\, changes in $Z$ can make significant differences in Main Sequence lifetimes, with differences of up to 0.4 dex in age between a 1 \msun\ star at Solar metallicity and one with $Z=0.001$, with the metal-poor stars having shorter lifetimes. Above 10 \msun\, the differences in $\log(\mathrm{age})$ stay small, generally no more than $\pm 0.05$ dex. Similar trends are seen to result from changing \afe\ at Solar $Z$ (middle panel). Below 10 \msun, differences in age can change by up to 0.4 dex between \afe$=-0.2$ and \afe$=+0.6$. Above 10 \msun, the differences are almost all less than $\pm 0.05$ dex. It should be noted that the $\alpha$-enhanced models, which have a correspondingly lower iron mass fraction, show shorter lifetimes. This is akin to the behaviour of the low $Z$, Solar-scaled models, emphasising that iron is the key element determining the opacity of the stellar envelope and hence luminosity, temperature and lifetime of the stars.

In the low metallicity scenario (lower-right panel), the differences remain largest at lower masses, spanning 0.1 dex at 1 \msun and decreasing close to zero for stars with an initial mass greater than 10 \msun. Thus it can be concluded that (assuming a fixed $Z$) the $\alpha$-enhanced massive stars that one would expect to find in the distant Universe have very similar Main Sequence lifetimes to those of a Solar-scaled composition. While still not accounting for binary evolution, this implies that CCSN delay time predictions are unlikely to be affected.

From a population synthesis perspective, another key property is the effective temperature of the stars, as this will determine the colour of the integrated light of the stellar population. As discussed above when considering the HR evolution tracks, changing the composition of the star can noticeably shift the effective temperature of the stellar models.

Fig~\ref{fig:TTAMS} is equivalent to Fig~\ref{fig:ATAMS}, but shows the effective temperature at the end of the Main Sequence rather than the stellar age, both in absolute values (left-hand panels) and relative values (right-hand panels). Below $\sim 40$ \msun\ ($\log(M/$\msun$=1.6$), a relatively smooth variation with composition change is seen. Both low $Z$ and $\alpha$-enhanced compositions (i.e. models with decreased iron content) show higher effective temperatures at the end of the Main Sequence than their Solar and Solar-scaled equivalents. At $Z=0.020$, these variations are around 0.1 to 0.2 dex in magnitude below $\sim 40$ \msun\ ($\log(M/$\msun$=1.6$). At $Z=0.002$ (lower panel of Fig~\ref{fig:TTAMS}) the variation in $T_{\mathrm{eff}}$ with $\alpha$-enhancement is less than 0.1 dex for all models less than $\sim$ 55 \msun ($\log(M/$\msun$=1.65$). 

At higher masses, the relation between composition and TAMS effective temperature changes noticeably. This is due to the sensitivity of massive star evolution to mass loss. If mass loss is comparatively low, a massive star would evolve to cooler temperatures, becoming a red supergiant (RSG). If mass-loss rates are higher, then the hydrogen-rich outer layers may be lost rapidly, exposing helium-rich material, and the star instead evolves towards higher effective temperatures at the end of its Main Sequence life, becoming a Wolf-Rayet (WR) or helium star. This dichotomy in Main Sequence evolution explains the sudden jump in TAMS effective temperature seen in Fig~\ref{fig:TTAMS} at high $Z$. Similar (but opposite) trends and behaviours can be seen looking at the stellar radius at the end of the Main Sequence, shown in the Supplementary Material (Fig~\ref{fig:RTAMS}). Metal-poor and $\alpha$-enhanced models show smaller radii than Solar and Solar-scaled models at the end of the Main Sequence. This change in physical size has consequences for binary stellar evolution, which will be discussed in Section~\ref{sec:bins}. The maximum difference in TAMS effective temperatures between Solar-scaled and $\alpha$-enhanced is $\sim$ 0.4 dex at $Z=0.020$ and $\sim$ 0.15 dex at $Z=0.002$.

\section{Post-Main Sequence evolution}\label{sec:PMS}

\subsection{Post-Main Sequence lifetimes}
\label{sec:PMS_life}

\begin{figure}
    \centering
    \includegraphics[width=0.98\linewidth]{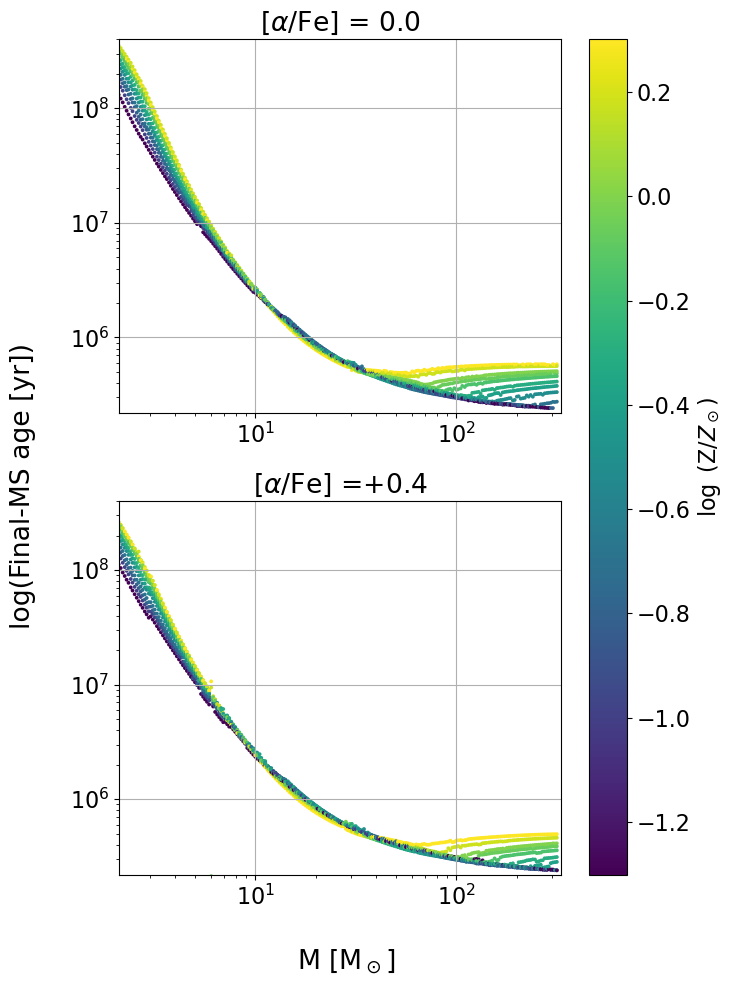}
    \caption{Post-Main Sequence lifetimes as a function of initial stellar mass. The upper panel shows the variation in lifetime at \afe$=0$, while the lower panel shows the same for \afe$=0.4$. The points are coloured in a logarithmic scale according to their total metallicity relative to `Solar', (i.e. $Z=0.020$). The variations are largest at both the low-mass ($M\lesssim5\,\mathrm{M}_\odot$) and high-mass ($M\gtrsim80\,\mathrm{M}_\odot$) ends, when AGB and massive star wind-mass loss dominate the post-Main Sequence evolution.}
    \label{fig:PostMS}
\end{figure}

Evolution beyond the Main Sequence is considerably shorter in terms of time, but has a significant impact on the ultimate outcome of stellar evolution, determining quantities as the masses of white dwarfs, the frequency of types of supernovae and their delay times, and the quantity of material expelled into the interstellar medium by stellar winds, among others. Fig~\ref{fig:PostMS} presents the Post-Main Sequence evolution timescale for Solar-scaled compositions (upper panel) and compositions with \afe$=+0.4$ (lower panel). The timescale is determined by subtracting the TAMS age derived earlier from the age at which the model has terminated or reached the top of the white dwarf cooling track, whichever is the soonest. This evolutionary timescale is compared across all values of $Z$ at \afe$=0$ (upper panel) and \afe$=+0.4$ (lower panel). For models with a $5\lesssim\mathrm{M}/M_\odot\lesssim70$, the post-Main Sequence lifetime is largely insensitive to changes in $Z$ or \afe. This is similar to what happens on the Main Sequence (left-hand panels of Fig~\ref{fig:ATAMS}) where stars in this mass range have very similar ages regardless of composition. At lower masses, the metal-rich stars show longer post-MS lifetimes, with the magnitude of the difference growing as the mass decreases, This is a result of the high mass-loss rates encountered on the AGB, which are heavily metallicity dependent. With a very high mass-loss rate, the total mass of the star decreases noticeably, thus extending its lifetime due to lower temperatures and slower nuclear reaction rates. A similar argument can be made in the case of the massive stars above 70\,\msun, where the high mass-loss rates lead to rapid stripping of the hydrogen-rich envelope. With higher metallicities and higher mass-loss rates, the surviving mass of the star is lower, prolonging its post-Main Sequence evolution.

At \afe$=+0.4$, the range of variation between the highest and lowest metallicities is reduced. This is a reasonable outcome, since the mass-loss rates are tied to the iron abundance and the post-MS evolution is highly sensitive to mass loss. The high-$Z$, $\alpha$-enhanced models have a lower iron abundance and thus evolve on a timescale equivalent to a model with a Solar-scaled mixture at a lower total $Z$. The shortest post-MS lifetimes appear very similar since the absolute change in iron abundance at $Z=1\times10^{-5}$ between Solar-scaled and $\alpha$-enhanced mixtures is negligible.

\subsection{Blue loops}
\label{sec:BLoops}

\begin{figure}
    \centering
    \includegraphics[width=0.485\textwidth]{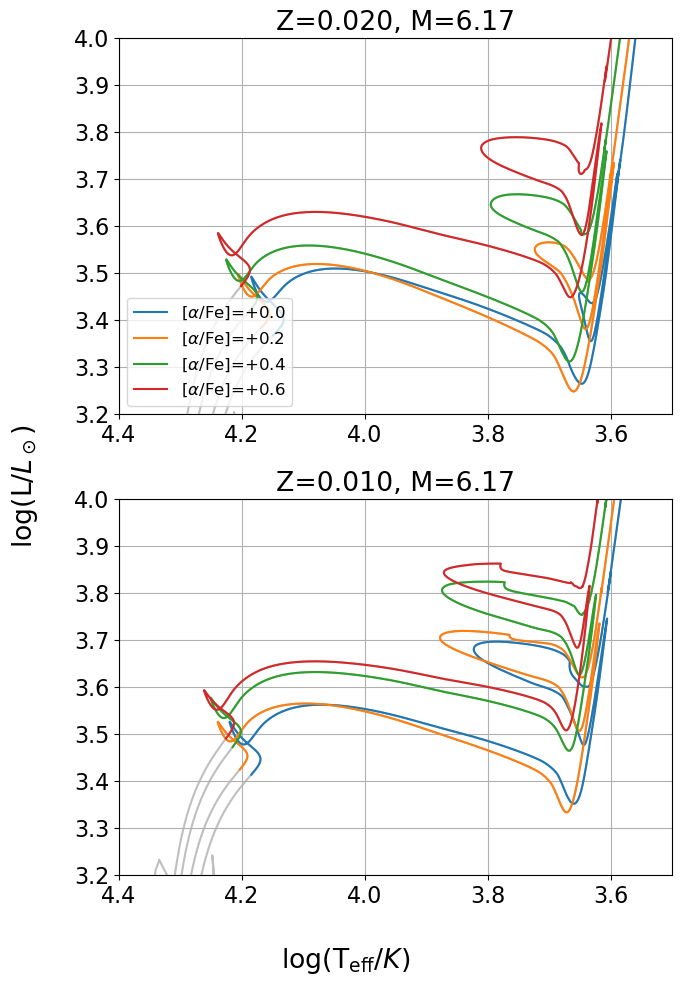}
    \caption{Evolution tracks highlighting the blue loops in 6.17\,\msun\ models at $Z=0.020$ (upper panel) and $Z=0.010$ (lower panel) for different values of \afe. The tracks are coloured from the point at which a helium-rich core has formed. A clear trend of bigger, brighter loops is seen as $Z$ decreases and/or \afe\ increases.}
    \label{fig:BLoops}
\end{figure}

Blue loops refer to the evolution from red to blue and back again seen in intermediate mass stars  ($4\lesssim\mathrm{M}/M_\odot\lesssim10$) during the helium-burning phase between the RGB and the AGB. The nature of the loop is sensitive to the opacity, helium abundance and mean molecular weight of the hydrogen burning shell \citep[][]{Walmswell15}. The behaviour of the blue loops has important consequences for understanding the population of Cepheid variables \citep[e.g.][]{Espinoza24}. The presence of blue loops is required to explain the population of Cepheids and thus are a necessary feature that 1-dimensional stellar evolution models must reproduce. The location and extent of the blue loops has some constraining power on assumptions regarding convection and stellar rotation \citep[e.g.][]{Johnston24}. Fig~\ref{fig:BLoops} shows a selection of evolution tracks for 7.94\,\msun\ (upper panels) and 6.17\,\msun\ (lower panels) stars at $Z=0.020$ (left-hand panels) and $Z=0.010$ (right-hand panels), with \afe\ variations shown in different colours. The general trend shows that $\alpha$-enhancement moves these loops to higher luminosities and extending to higher temperatures. This is similar to the manner in which the loops to get more luminous and more extended at lower total metallicity (as seen in the right-hand panels).

\subsection{Wolf-Rayet stars}\label{sec:WR}

\begin{figure}
    \centering
    \includegraphics[width=0.485\textwidth]{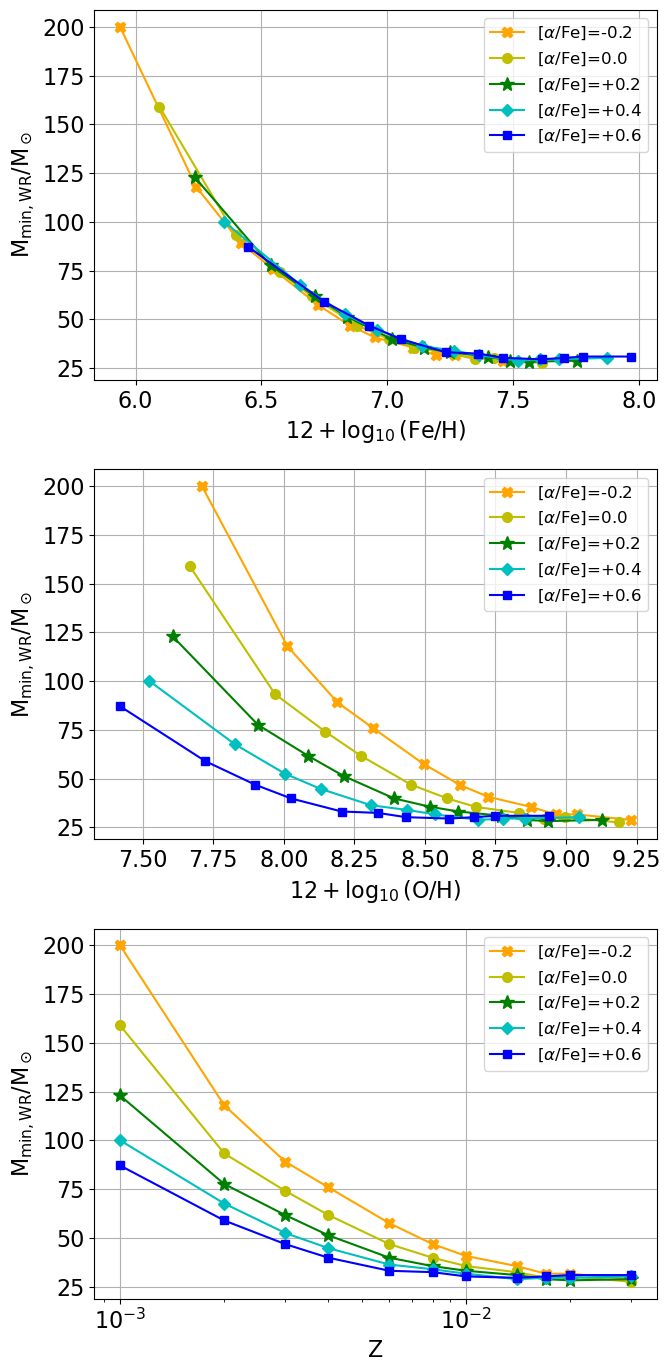}
    \caption{Minimum initial mass to form a Wolf-Rayet star as a function of [Fe/H] (upper panel), [O/H] (centre panel), and metallicity mass fraction $Z$ (lower panel), all coloured according to \afe. A strong correlation with \feh\ is observed.}
    \label{fig:Min_WR}
\end{figure}

Wolf-Rayet (WR) stars \citep[][]{WolfRayet1867} are hot, luminous stars, whose spectra are characterised by strong emission lines arising from strong, optically thick stellar winds. Classical WRs refer to those which lose all of their hydrogen envelope to their strong winds and present a helium- (or carbon-oxygen-) dominated atmosphere. These stars are a major source of ionising radiation in stellar populations and are a proposed source of long gamma ray bursts \citep[LGRBs, e.g.][]{Crowther07,Woosley07,Song23}. Fig.~\ref{fig:Min_WR} highlights the minimum mass necessary for stars to form a classical WR star, as a function of iron abundance (upper panel), oxygen abundance (centre panel) and total metallicity (lower panel). This was determined by identifying the lowest mass model retaining a surface hydrogen abundance less than 1 per cent at the end of its evolution. It is readily apparent that iron abundance is the dominant factor determining the minimum mass, with all curves in the upper panel following the same trajectory, from $\sim30$ close to Solar, to 200\,\msun\ at very low iron abundance. Considering a fixed value of \oxh\ or $Z$, large variations in the minimum mass can be seen. For example, at \oxh$=8.0$, the minimum initial mass that forms a WR star lies between 40\,\msun\ and 120\,\msun. As a consequence, relying on oxygen abundances to infer the metallicity of a stellar population will lead to a large uncertainty on the number of WR stars and thus significant uncertainty in the expected amount of ionising flux (He\,{\sc{ii}} in particular) and the number of long GRB progenitors in the population.

\subsection{Initial-to-Final mass relation}

\subsubsection{Low-mass stars: White Dwarf masses}

\begin{figure}
    \centering
    \includegraphics[width=0.48\textwidth]{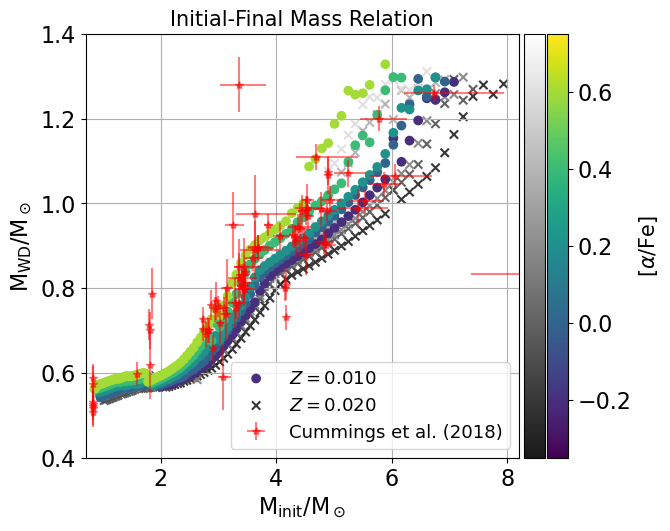}
    \caption{Initial-to-final mass relation for white dwarfs as a function of $Z$ and \afe. Models with a metallicity of $Z=0.010$ are shown as coloured circles while those with a metallicity of $Z=0.020$ are shown as greyscale crosses. Also shown as red stars are the values derived by \protect\citet{Cummings18} from an observational sample of 80 nearby white dwarfs.}
    \label{fig:WD_IFMR}
\end{figure}

The initial-to-final mass relation (IMFR) is an important quantity for understanding the population of white dwarfs. Many theoretical and observational determinations of the relation have been made \citep[e.g.][]{Weidemann83,1995A&A...297..727B,MillerBertolami16,Cunningham24}. Fig~\ref{fig:WD_IFMR} shows the initial-to-final mass relation derived from our models for stars below $\sim8$\msun and how it vaires with $Z$ and \afe. The comparison is done at both solar ($Z=0.020$, greyscale crosses) and half-Solar ($Z=0.010$, coloured circles) metallicity mass fractions. For initial masses of $M/\mathrm{M}_\odot\lesssim3$, changes in composition have a relatively minor impact on the final mass. Above this point, the relation shows the same characteristic shape for all compositions, but is noticeably steeper when $Z$ is lower and/or \afe\ is higher. For example, taking a star with an initial mass of 5\,\msun\ and $Z=0.020$, the final white dwarf mass could vary between 0.9 and 1.1\,\msun, depending on the value of \afe. A consequence of a steeper IMFR is that the proportion of massive white dwarfs that is expected in a stellar population is higher, since they form from less massive stars, which are more numerous.

The initial and final masses of determined by \citet{Cummings18} for a sample of 80 nearby white dwarfs are indicated by the red stars. Comparing to the \bpass\ models, the bulk of the population lie along the theoretical relations plotted. The scatter in the observations is comparable to the spread in the relation arising from changes in \afe. This suggests that \afe\ -- alongside $Z$ and binary interactions -- is an important parameter to consider when inferring the initial mass of a white dwarf progenitor. At this point, the reader should be reminded that the models presented here only consider single-star evolution and binary interactions (mass transfer and stellar mergers in particular) will add additional uncertainty and scatter to these relations, as discussed in \citet{2017PASA...34...58E}.

\begin{figure*}
    \centering
    \includegraphics[width=0.98\textwidth]{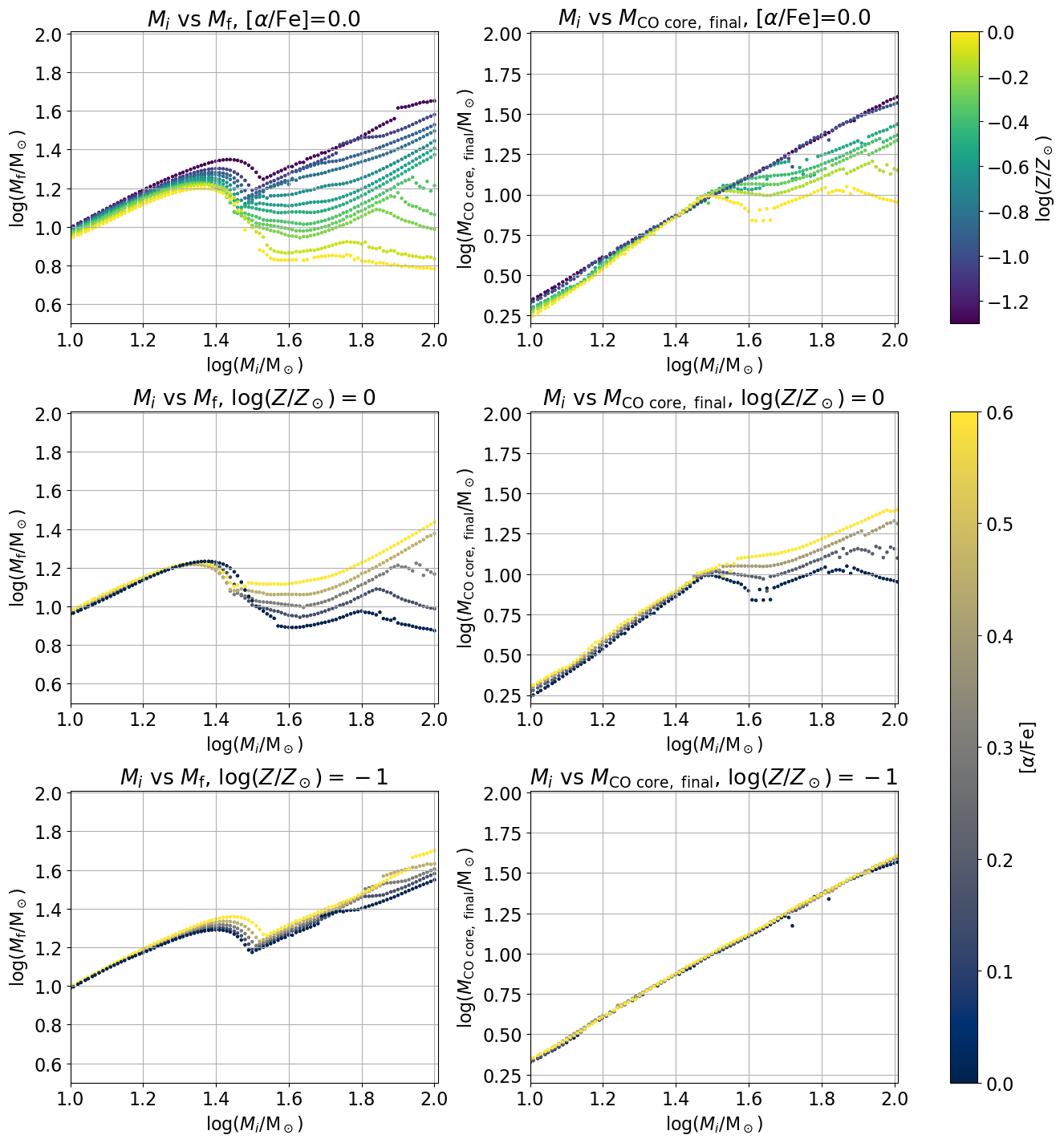}
    \caption{Left panels: Initial to final mass relation for various compositions. Upper panel: Solar-scaled compositions, varying Z. Centre panel: Z=Z$_\odot$=0.020, varying \afe. Lower panel: Z/Z$_\odot$=0.1, varying \afe. Right-hand panels show the CO core mass as a function of initial mass, with each panel showing the same variation in composition as the corresponding left-hand panel.}
    \label{fig:MFINA}
\end{figure*}

\subsubsection{Massive stars}

The left-hand panels of Fig~\ref{fig:MFINA} show the relationship between the initial stellar mass and the mass at the end of the calculation for stars with an initial mass greater than 10 \msun. Below $\sim$25 \msun\ ($\log(M/$\msun$)=1.4$), the initial-to-final mass ratio behaves in a simple and monotonic manner, with metal-rich stars having lower final masses, owing to their higher mass-loss rates. At higher masses, the lowest metallicity models show a systematic downward shift in the final mass of about 0.2 dex. This arises from the transition from the use of the \citet{deJager88} mass-loss rates for cool Main Sequence stars to the \citet{Vink01} rates for stars with effective temperatures greater than $25\,000\,$K. At higher metallicities, the final mass shows a second strong decrease, associated with the transition to \citet{Nugis00} WR mass-loss rates as helium-rich, processed material becomes exposed at the stellar surface.
It is important to note that `final' mass refers to the mass of the star at the end of the calculation (i.e., after core carbon burning which we assume to be just prior to core-collapse) rather than the mass of the post-explosion remnant.

This behaviour is qualitatively similar for Solar-scaled compositions with varied $Z$ (upper left panel) and  $Z=0.020$ compositions with varied $\alpha$-enhancements (centre left panel). The metal-rich, $\alpha$-enhanced stars show higher final masses. This is consistent with the fact that they have lower iron abundances than the equivalent Solar-scaled models. It is the iron opacity which has the strongest impact on the evolution of the massive stars and their mass-loss rates. The same trends can be seen with changes in mass and composition as in the upper left panel, albeit that the final masses of the $\alpha$-enhanced models are comparable to those with a lower $Z$ and a Solar-scaled composition. The differences in final mass arising from changing \afe\ at 0.1 $Z_\odot$ (lower left panel) are much smaller, since the iron mass fraction is already rather small.

The mass at which this second decrease in the initial-to-final mass relation happens appears to be composition dependent. Decreasing the abundance of iron and metals in general changes the opacity and radius of the star, with metal rich stars having larger radii. As a result, their mass-loss rates are higher and they lose their hydrogen rich envelopes more rapidly than their metal-poor counterparts. This leads them to them evolving into WR stars, which have strong optically thick winds, thus exacerbating their mass loss. At lower metallicities, this transition requires a higher initial mass to manifest, where the luminosity becomes great enough to remove the hydrogen envelope and allow WR evolution to take place. This agrees with the minimum masses required to form WR stars, as discussed in Section~\ref{sec:WR} above.

Consider, for example, the lower right panel of Fig~\ref{fig:hr50100}. The 100 \msun\ model at Solar metallicity evolves to lower luminosity but higher temperature as the extreme mass loss leads to it forming a WR star, while the lower metallicity models with their lower Main Sequence mass-loss rates initially evolve to cooler temperatures, becoming RSGs, as their outer layers are not removed rapidly enough to expose the helium-rich material before core hydrogen exhaustion.

\subsection{Supernova progenitors}
\label{sec:sNE}

SN progenitor stars are valuable tools to constrain the physics and evolution of massive stars. RSGs, which result in a Type II SN have a very narrow range of surface temperatures, with their luminosity being directly related with their stellar mass. Thus, the maximum inferred brightness of a SN progenitor provides a constraint on the most massive stars which become RSGs. One outstanding challenge in this area is the discrepancy between the most luminous RSGs predicted from stellar modelling and those found in observations, the so-called `missing' high-mass stars \citep[see e.g., the review of][]{Smartt15}. Single-star stellar evolution models predict that RSGs can be formed by stars with initial mass of up to $\sim$ 30 \msun, while no confirmed progenitors have been observed with inferred initial masses larger than $\sim$ 20 \msun.

\begin{figure}
    \centering
    \includegraphics[width=0.476\textwidth]{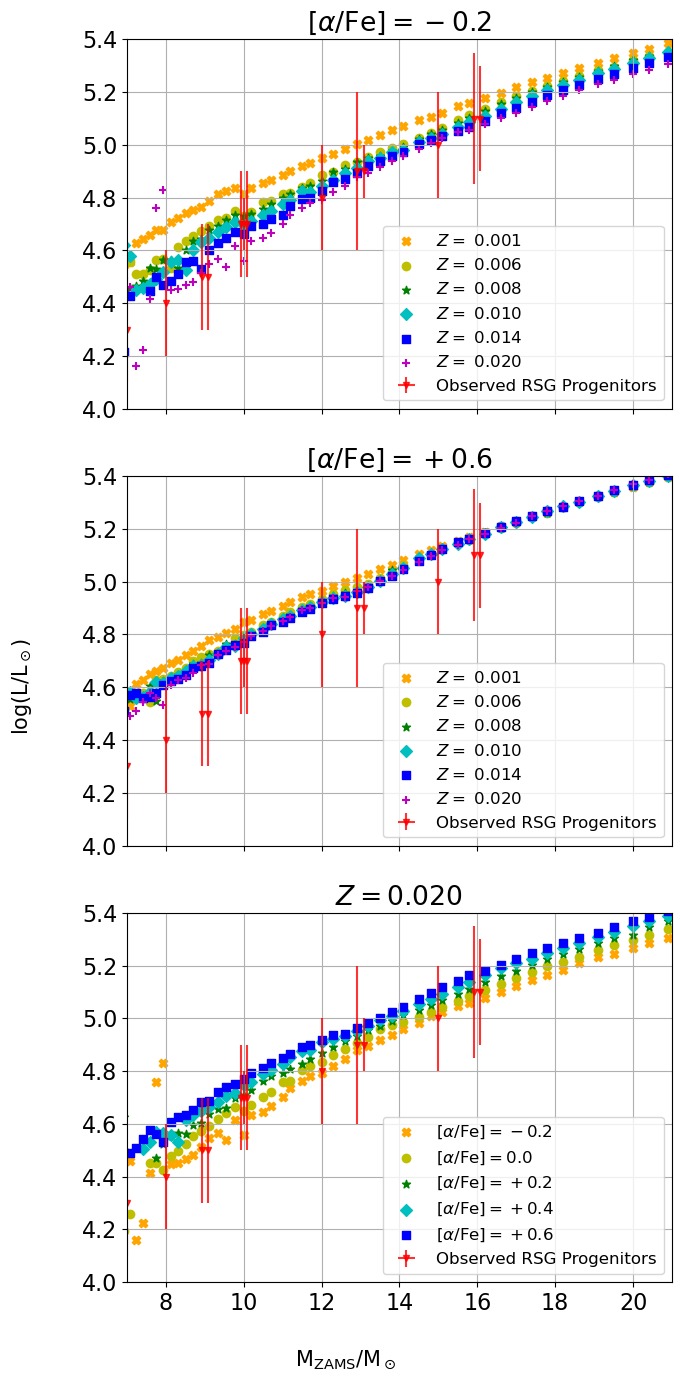}
    \caption{Initial mass -- final luminosity relation for RSGs, compared to observational data from \protect\citet{Smartt15} and \protect\citet{ONeill21} (red triangles). Upper panel: $\alpha$-depleted, iron-rich compositions with $Z$ varied from 0.001 to 0.020. Centre panel: $\alpha$-enhanced, iron-poor compositions with $Z$ varied from 0.001 to 0.020. Lower panel: Bulk metallicity of $Z=0.020$, with \afe\ varying from $-0.2$ to $+0.6$.}
    \label{fig:sNE_ML}
\end{figure}

Fig~\ref{fig:sNE_ML} shows the final RSG luminosity of the \bpass\ models as a function of initial stellar mass for a wide range of metallicities. The cyan circles indicate models with Solar-scaled composition, while the magenta diamonds represent models with an $\alpha$-enhancement of \afe$=+0.6$. Masses and luminosities of confirmed SN progenitors are plotted as red triangles. Values for the progenitors are taken from Table 1 of \citet{Smartt15}, with the exception of SN 2008bk which uses the revised values from \citet{ONeill21}. A small offset is seen between the Solar-scaled and $\alpha$-enhanced luminosities, which grows with increasing $Z$. The $\alpha$-enhanced models are preferentially more luminous. Thus, if one were to assume a star were $\alpha$-enhanced, rather than Solar-scaled in composition, the inferred stellar mass would be lower, which does not help to address the lack of massive RSGs. The uncertainties in the observed luminosities are greater than the differences produced by composition changes, so considerations of $\alpha$ abundance variations are unlikely to play a major role in the mass determination of RSGs in the context of SN progenitors.

Another aspect of the observed RSG SN progenitor samples is that there is an excess of low luminosity progenitors \citep{2011MNRAS.417.1417F}. Models yield a minimum luminosity of a RSG SN progenitor because, below a certain threshold, less massive stars eventually go through second dredge-up, which leads to the luminosities of stars increasing by nearly an order of magnitude \citep[e.g.][]{Eldridge04b,2007MNRAS.376L..52E}. Previous {\sc{stars}}/\bpass\ stellar models have struggled to obtain a SN progenitor luminosity $\log(L/L_{\odot}) \la 4.5$. This is in tension with several observed pre-SN progenitor luminosity measurements and limits, as detailed by \citet{2011MNRAS.417.1417F}. However in the new model grid presented here, we see that low \afe\ and Solar metallicity models do show a decrease in luminosity compared to the standard Solar mixture. This would be the expected composition of most SNe with observed progenitors, which are limited to the very local Universe. More detailed measurements of the composition of the host galaxies of the SN progenitors, and in particular measurements of multiple elemental abundances, would be required to test this hypothesis.

A broader look at the final HRD locations across the entire range of massive stars -- and the impact of \afe\ variation on this -- is shown in the Supplementary Material (Section~\ref{app:SNe}). To summarise, $\alpha$-enhancement tends to increase both the maximum luminosity and mass of RSGs, before the population transitions to blue supergiants (BSGs). This is consistent with the interpretation that the $\alpha$-enhanced models are iron-poor, therefore have lower mass-loss rates, and thus are able to maintain their hydrogen envelopes, remaining as RSGs.

An additional detail that can be determined using these models is the minimum initial mass required for a Fe-core collapse SN using our prescription. Here we select the minimum mass model (at each composition) that has a CO-core mass above 1.4~M$_{\odot}$ and has experienced core carbon burning. In several composition grids we sometimes accept a slightly lower mass that is above 1.39~M$_{\odot}$ but has a very high central density above $10^8 {\,\rm g\,cm^{-3}}$. With these constraints we are attempting to exclude stellar models that have gone through, or will go through, second dredge-up before core collapse. These models might be expected to experience electron capture SNe in an oxygen-neon core, and so may look quite different to a standard SN. The mass limits derived are also consistent with the more detailed models of \citet{2015ApJ...810...34W}.

\begin{figure}
    \centering
    \includegraphics[width=0.476\textwidth]{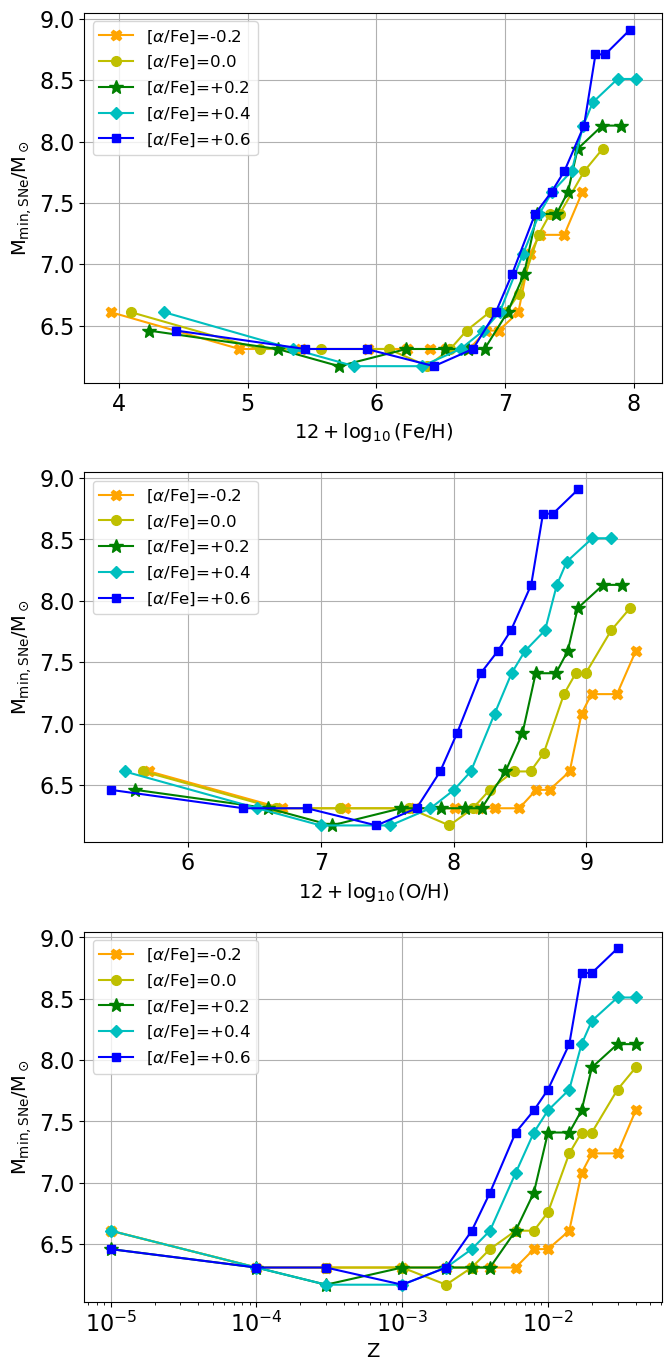}
    \caption{Minimum mass progenitor of a Fe-core collapse SN as a function of [Fe/H] (upper panel), [O/H] (centre panel), and metallicity mass fraction $Z$ (lower panel), all coloured according to \afe.}
    \label{fig:minmasses}
\end{figure}

We show how the minimum initial mass for SN varies with {12~+~log$_{10}$(Fe/H)} and {12~+~log$_{10}$(O/H)} in Fig~\ref{fig:minmasses}. As can be seen in the lower panel, there is a minimum initial mass at low bulk metallicities of about 6.2~M$_{\odot}$. This increases slightly at the lowest metallicities, but can be as high as 9~M$_{\odot}$ at our highest metallicities. As a result, the number of SNe from a metal-rich population would be substantially lower. Comparing the different compositions, it is clear that models with the same iron abundance are very consistent with one another except at the highest abundances. However when considering fixed {12~+~log(O/H)} abundance we see that the range of possible minimum initial masses permitted for core collapse varies from 6.31 to 7.7~M$_{\odot}$ at {12~+~log(O/H)}=8.5, and from 6.46 to 8.71~M$_{\odot}$ at {12~+~log(O/H)}=8.7. This is a significant difference, which will come into play in comparison to observational samples, which often use {12~+~log(O/H)} as a proxy for bulk metallicity in order to to calculate the SN rate. For example, \citet{2023ApJ...955L..29P} recently reported a significant change in the CCSNe rate over a moderate change in metallicity, in a study in which {12~+~log(O/H)} was used as the measurement of metallicity. 

Using the models presented here, we can see here that such a measurement is highly sensitive to $\alpha$-enhancement and has very little to do with the range of masses that are expected to end their lives in CCSNe. Using a simple IMF calculation we can estimate that the difference in number of SNe expected due to changing metallicity at \afe=+0.6 is a factor of 1.3 from the highest to lowest metallicity. While for \afe=-0.2 it is a factor of 1.6. This is not a full explanation for the order of magnitude difference found by \citet{2023ApJ...955L..29P}, but does indicate that when the iron abundance is unknown, it is difficult to precisely calculate the expected SN rate.

While the lower end of the SN progenitor mass range is of interest, the more massive SN progenitors which form black holes are also of significant interest given the growing number of gravitational wave (GW) transient detections \citep[e.g.][]{2023PhRvX..13d1039A}. The masses of compact remnants can be approximated from the CO core masses using prescriptions such as those in \citet{2012ApJ...749...91F}. Thus it is useful to understand the effect of metallicity and \afe\ on the CO core masses which we show in the right hand panels of Figure \ref{fig:MFINA}. We see that the CO core masses are significantly affected by metallicity at masses above about 30~M$_{\odot}$. However the \afe\ composition only has a significant impact on the CO core masses at solar and moderate metallicities. At lower metallicities the core masses all tend towards the same values, driven by the sizes of the convective cores in a star that has lost little of its initial mass. Thus we can conclude that determining the correct \afe\ composition for metal-rich stars is important to estimate their remnant masses accurately. Given we only evolve our models up to core carbon burning these results should only be considered indicative.

In earlier \bpass\ model releases we have also estimated the expected compact remnant masses using a simple calculation of how much mass is removed from the star if we inject $10^{44}$~J of energy into the envelope \citep{Eldridge04b}. This provides estimated compact remnant masses that are similar is magnitude to other prescriptions \citep[e.g.][]{2023MNRAS.520.5724B}. We show these in Figure \ref{fig:SNe_M_rem} and the results confirm our above findings on the CO core mass. There appears to be more significant variation in the estimated remnant masses than the CO core masses. This suggests the core compactness at high metallicity may be affected by \afe\ composition to a significant degree.

Finally, it is important to recall that these massive stars are highly likely to have binary companions \citep[][]{2012Sci...337..444S}. As these stars will have very large radii, the possibility of binary interactions is high. This will reduce the fraction of massive RSGs reaching the end of their lives unaffected by a companion. In the era of large all-sky surveys, such as the Legacy Survey of Space and Time at the Vera Rubin Observatory, the population of very luminous RSG progenitors may provide constraints on the wide binary fraction of these massive stars. A fuller interpretation of the most massive SN progenitors will require calculation of large grids of binary stellar models and further consideration of very massive star mass loss rates. This is an area of ongoing effort (Byrne et al., in prep.).

\section{Fixed iron models}
\label{sec:fixFe}

While total metallicity $Z$ is the approach commonly used by theorists, since the composition of their stellar model can be precisely defined, observers typically rely on the strength and shape of spectral features to determine metallicity, often using just one or a small number of elements (e.g \feh, \oxh). For example, in galaxy observations, 12~+~log(O/H) is often used as the key metallicity tracer. This is typically derived from nebular emission lines, which are not necessarily representative of the metallicity or composition of the underlying stellar population.

To address this, we also examine how evolution is affected when we vary \afe, while keeping an observationally motivated metallicity indicator fixed, rather than a theoretical one. We consider models which have differing total $Z$ and \afe\ values, but an equivalent (or near equivalent) iron abundance, [Fe/H]. Given the discrete values of $Z$ and \afe\ at which the models have been calculated, an exact match on iron abundance is not possible, but we select models which are closely comparable. The selected compositions are shown in Table~\ref{tab:sameFe} alongside the mean and standard deviation of [Fe/H] for each sequence. 7 sequences were selected, ranging from 5 per cent to 100 per cent of the Solar value in iron.

Comparative HR diagrams for 20 \msun\ models are shown in the Supplementary MAterial (Section~\ref{app:samefe}). Here we provide a brief summary of the key findings. The differences in the evolution tracks are smaller than those arising from composition changes seen in Fig~\ref{fig:hr1020}. The variation in evolution tracks gets larger moving to higher metallicities. This provides a clear indication that the iron mass fraction is the biggest factor in determining the evolution of a given model, rather than changes in the bulk $Z$ or \afe\ specifically.

The differences in TAMS age and TAMS radius are typically small when the total metallicity is low. Considering Main Sequence lifetimes, the differences from Solar-scaled composition remain less than 0.1 dex up to and including 30 per cent solar. For models with more than 30 per cent of the Solar iron abundance, the differences continue to grow, particularly at the low mass end. In the case of 100 per cent Solar iron abundance, a Solar mass star with an $\alpha$-enhancement of 0.4 dex has a lifetime which is almost 0.2 dex shorter than the Solar composition equivalent.

The fact that some differences remain when using models with matched iron abundances indicates that while iron plays the dominant role in determining the course of the star's evolution, the relative abundance of other elements also contributes to the atmospheric parameters and evolution. As one would expect, the effect is more pronounced when the overall metallicity is higher, as a larger contrast in composition can be achieved. This is important to consider as many different models exist in the literature and each uses a different definition of Solar composition.

\begin{table*}
    \centering
    \caption{Combinations of \afe\ and $Z$ selected to have near-identical iron abundances, along with their means and standard deviations.}
    \begin{tabular}{c|c|c|c|c|c|c|c}
        & \afe=-0.2 & \afe=+0.0 & \afe=+0.2 & \afe=+0.4 & \afe=+0.6 & <$N_\mathrm{Fe}/N_{\mathrm{Fe}_\odot}$> & $\sigma(N_\mathrm{Fe}/N_{\mathrm{Fe}_\odot})$ \\\hline
    Fe1, $N_\mathrm{Fe}/N_{\mathrm{Fe}_\odot}$ $\simeq 0.05$  & ---            & $Z=$ 0.001     & ---            & $Z=$ 0.002     & $Z=$ 0.003     & 0.0494 & 0.0003 \\
    Fe2, $N_\mathrm{Fe}/N_{\mathrm{Fe}_\odot}$ $\simeq 0.10$  & ---            & $Z=$ 0.002     & $Z=$ 0.003     & $Z=$ 0.004     & $Z=$ 0.006     & 0.1007 & 0.0033 \\
    Fe3, $N_\mathrm{Fe}/N_{\mathrm{Fe}_\odot}$ $\simeq 0.20$  & $Z=$ 0.003     & $Z=$ 0.004     & $Z=$ 0.006     & $Z=$ 0.008     & ---            & 0.2009 & 0.0075 \\
    Fe4, $N_\mathrm{Fe}/N_{\mathrm{Fe}_\odot}$ $\simeq 0.30$  & ---            & $Z=$ 0.006     & $Z=$ 0.008     & ---            & $Z=$ 0.020     & 0.3061 & 0.0225 \\
    Fe5, $N_\mathrm{Fe}/N_{\mathrm{Fe}_\odot}$ $\simeq 0.50$  & $Z=$ 0.008     & $Z=$ 0.010     & $Z=$ 0.014     & $Z=$ 0.020     & $Z=$ 0.030     & 0.5066 & 0.0095 \\
    Fe6, $N_\mathrm{Fe}/N_{\mathrm{Fe}_\odot}$ $\simeq 0.70$  & $Z=$ 0.010     & $Z=$ 0.014     & $Z=$ 0.020     & $Z=$ 0.030     & $Z=$ 0.040     & 0.7046 & 0.0354 \\
    Fe7, $N_\mathrm{Fe}/N_{\mathrm{Fe}_\odot}$ $\simeq 1.00$  & $Z=$ 0.014     & $Z=$ 0.020     & $Z=$ 0.030     & $Z=$ 0.040     & ---            & 1.0060 & 0.0622 \\
    \end{tabular}
    \label{tab:sameFe}
\end{table*}

\section{Comparisons to other work}

\subsection{Observations}
Resolved stellar populations such as a stellar cluster provide an opportunity to test our models against observations. Here we select three nearby open clusters where individual stars have well defined atmospheric parameters (surface gravity and effective temperature) derived from spectroscopy. We consider a young open cluster: Matariki/Pleiades/M45\footnote{As a bright, naked-eye star cluster, this object has a large number of names attributed to it by cultures around the world. In the M\={a}ori tradition of Aotearoa New Zealand, it is known as Matariki \citep[][]{Matamua20,Harris13Matariki,Condliffe1919Matariki} and its appearance in the dawn sky marks the start of the New Year.}, a cluster of intermediate age, NGC~6811, and an old open cluster: M67/NGC~2682 \citep[][]{Messier1781,DreyerNGC}. By plotting isochrones (curves of models of equal age) alongside observational data, we can evaluate how well our models recover existing estimates of cluster ages and compare the uncertainties arising from changes in total metallicity to those from changes in \afe. A rigorous fitting of the isochrones to the observational data is carried out using the {\sc{hoki}} AgeWizard tool \citep[][]{HOKI,Stevance20AgeWizard}. This compares the density of observed points in the effective temperature--surface gravity plane to those of a \bpass\ simple stellar population over all ages to derive a probability distribution function and infer the most likely age of the cluster. Here we discuss some of the most important outcomes of the fitting, further detail and the probability distribution functions can be found in the Supplementary Material (Section~\ref{app:clusfit}).

The general findings is that using isochrones with a higher \afe\ leads to older age estimates, with a spread of around 0.4 dex in log(age) between models with \afe$=-0.2$ and \afe$=+0.6$, for the old and young stellar clusters. At intermediate ages, the change is smaller, with a 0.1 dex change in the best fitting age. Here we outline the results found for fitting each of these three clusters and show qualitatively how these changes in age and composition can give visually similar isochrones. The fitting is described in more detail in the Supplementary Material (Section~\ref{app:clusfit}).


\subsubsection{Young cluster: Matariki/Pleiades/M45}
M45 is a bright, nearby open cluster with age estimates of 80-120 Myr \citep[e.g.][]{Mermilliod81,Gaia18b,Heyl22} and a metallicity similar to Solar. As a nearby cluster, derivation of the atmospheric parameters of individual stars is relatively straightforward with good quality spectroscopy. An observational sample was obtained from the APOGEE/SDSS public data release \citep[][]{Majewski17APOGEE,Abdurrouf22SDSS17}. Stars within 30 arcminutes of the centre of the cluster were initially selected. To remove interlopers, stars which had a parallax more than 1 standard deviation from the mean value were removed, along with stars without surface gravity determinations, leaving 317 stars.

\begin{figure}
    \centering
    \includegraphics[width=0.48\textwidth]{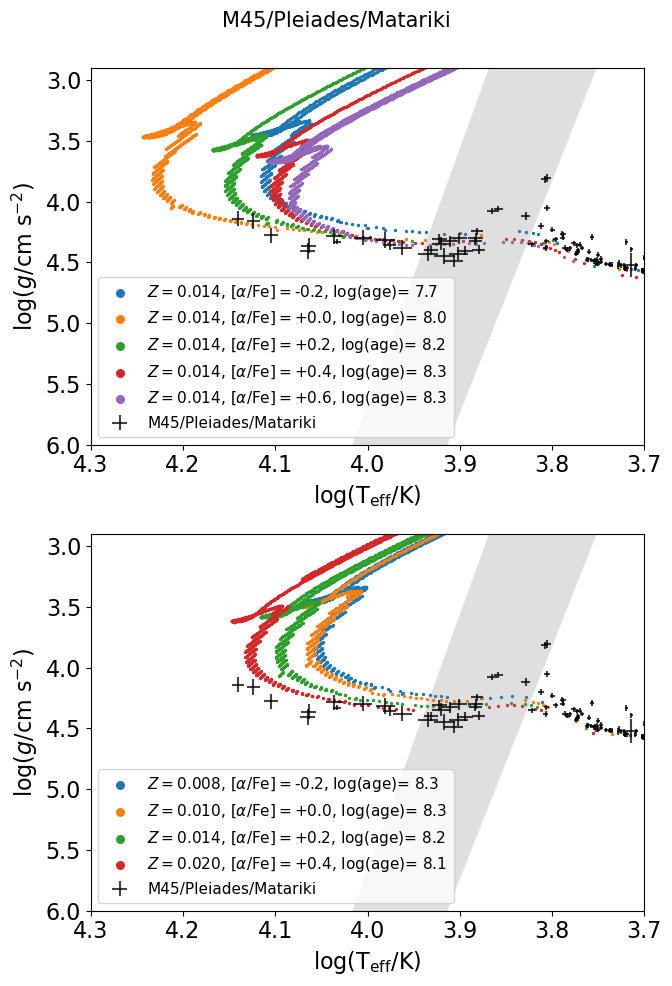}
    \caption{Kiel diagram showing observationally determined spectroscopic parameters for M45/Pleiades (black crosses) from APOGEE/SDSS compared to our isochrones of various ages, metallicities and \afe\ values. Upper panel: Fixed composition with varied age, Second panel: Fixed age, varied bulk metallicity, Z, Third panel: Fixed Z and age, varied \afe, Lower panel: Fixed Z, varied age and \afe. Isochrones are defined by plotting all points in stellar evolution models within 0.02 dex of the specified log(age). The shaded region indicates the location of the classical instability strip.}
    \label{fig:Pleiades}
\end{figure}

Fig~\ref{fig:Pleiades} shows the location of these stars in a log($g$)-T$_\mathrm{eff}$, or Kiel diagram, alongside isochrones produced from our models at different ages, bulk metallicities and $\alpha$-enhancement values. Our plotted isochrones are smeared out by 0.01 dex in log(age) to provide sufficient model points for the curves to be easily seen. The upper panel shows isochrones with fixed $Z$ and varied \afe, while the lower panel shows a selection of $Z$ and \afe\ combinations where the iron content is approximately the same, as per Table~\ref{tab:alpha_abundances}. The plotted ages correspond closely to the median values determined from fitting with Age Wizard.

The deviation seen between the models and the observations at temperatures between log(T) of 3.8 and 3.9 can be attributed to stars here lying in the classical pulsation instability strip, indicated by the shaded region. As such, the surface gravity measurements may be less reliable. Using $Z=0.014$, the $\alpha$-depleted isochrone shows a significantly younger stellar population is needed to match the data. The remainder show a variation of $0.3$\,dex in log(age), with the $\alpha$-enhanced isochrones preferring an older age to fit the data. The spread in age estimates, the broad probability distribution in the age fitting and the variety of isochrone shapes seen here serve to highlight the difficulty in aging a young cluster with a lack of post-Main Sequence objects. A log(age) of 8.2 or 8.3 is slightly older than the literature values for the age of the cluster, but visual inspection of Fig~\ref{fig:Pleiades} shows that these isochrones don't necessarily capture all of the hottest stars, and a slightly younger age may match better. Age Wizard is designed for fitting observations of stellar clusters to models of binary stellar populations, and as such may not be best suited to handling single star population which form a neat line, rather than an extended region.

\subsubsection{Intermediate age: NGC~6811}

\begin{figure}
    \centering
    \includegraphics[width=0.48\textwidth]{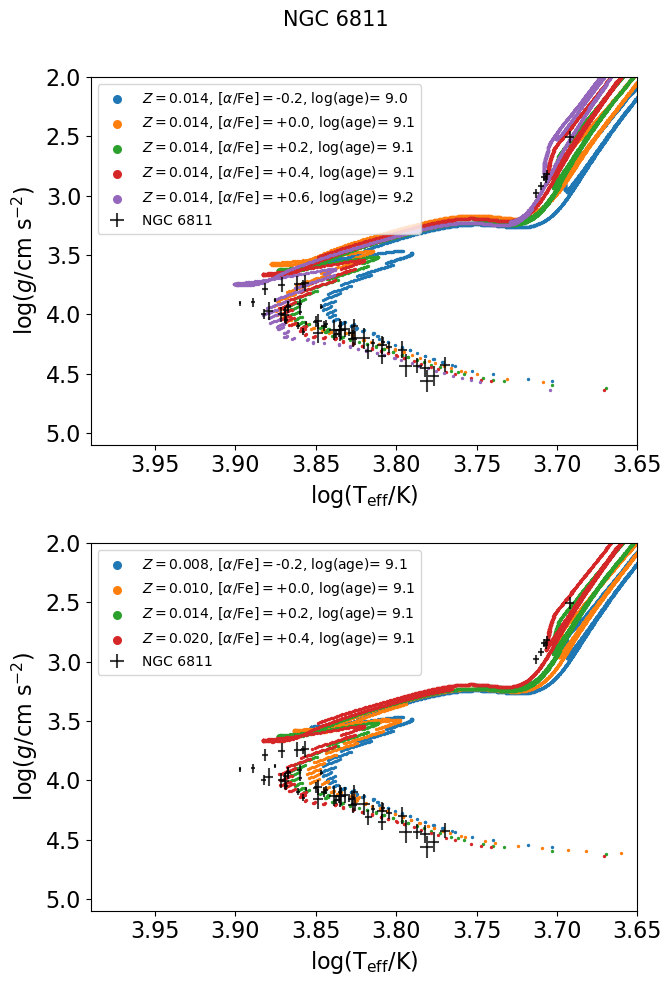}
    \caption{As Fig~\protect\ref{fig:Pleiades}, but with LAMOST observations of NGC~6811.}
    \label{fig:NGC6811}
\end{figure}

NGC~6811 is an open cluster with a moderate age, with literature ages of around 1 Gyr and a metallicity between Solar and 60 per cent Solar \citep[e.g.][]{Mills05,Janes13,MolendaZakowicz14}. Surface gravities and effective temperatures for stars in this cluster were taken from spectroscopic data in the LAMOST survey \citep[][]{Luo15LAMOST}, with cluster membership as assigned in the data tables of \citet{Fu22LAMOST} who combined the LAMOST observations with Gaia proper motions to do so. As the LAMOST data is cross-matched with Gaia DR2, which had very limited data on variable stars, the selected stars are not likely to be contaminated by classical pulsators as in M45 above. A total of 59 stars are included in the dataset, once outliers with large errors in the atmospheric parameters were removed. We compare the observations to some of our theoretical isochrones in Fig~\ref{fig:NGC6811}, following a similar approach to Fig~\ref{fig:Pleiades}.

For a fixed metallicity of $Z=0.014$, we once again see a preference for older ages as \afe\ increases. A difference in log(age) of 0.2 dex is seen, although it is only the most extreme values of \afe\ that deviate from log(age)$=9.1$ as the median fit. This provides a good match to existing literature estimates for the age of the cluster. At this age ($\sim1$\,Gyr), the increased temperatures from higher \afe are almost equally balanced by the shorter evolutionary timescales, yielding very similar turn-off points. Keeping the iron abundance fixed (lower panel), there is good agreement between all models. The presence of a clear turn-off and some post-Main Sequence objects provide much stronger constraining power, as evidenced by the sharp peak in the probability distribution of the fit.

\subsubsection{Old open cluster: M67/NGC~2682}
\begin{figure}
    \centering
    \includegraphics[width=0.48\textwidth]{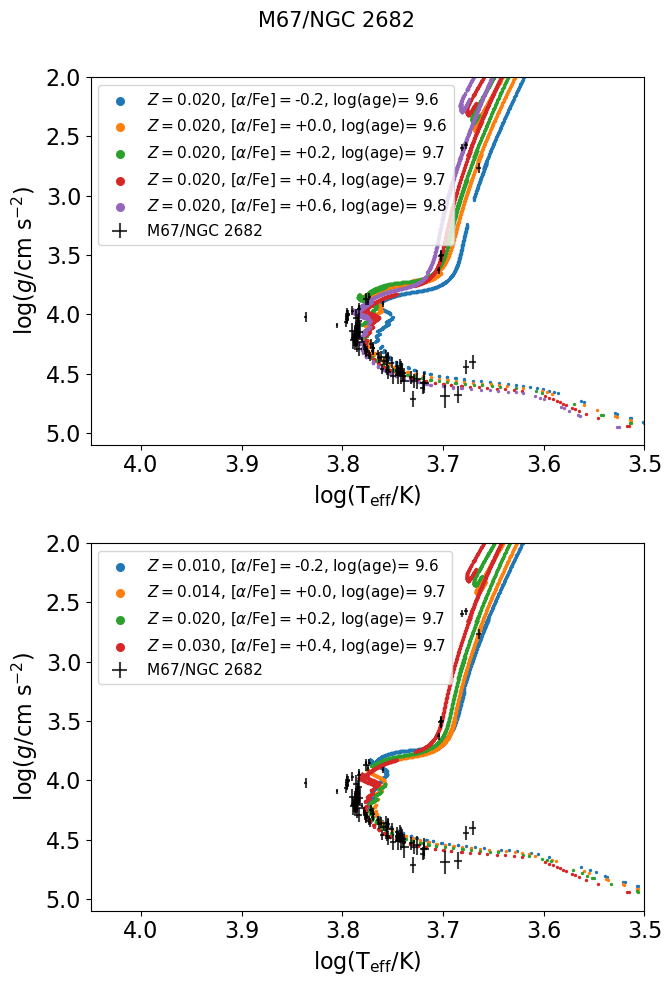}
    \caption{As Fig~\protect\ref{fig:Pleiades}, but with LAMOST observations of M67/NGC~2682.}
    \label{fig:M67}
\end{figure}

Thirdly, we consider M67/NGC~2682, one of the oldest nearby open clusters. Literature values for the age of the cluster are typically around 4 Gyr \citep[e.g.][]{Richer98,Hurley05,Kharchenko13} with a metallicity very close to Solar \citep[e.g.][]{Onehag11}. Once again we take observational parameters for individual stars from LAMOST. A total of 100 stars are included. This sample of stars has a well defined Main Sequence turn-off and a number of subgiant and red giant branch stars, which are useful for constraining the model choices, with the median isochrone fits plotted in Fig~\ref{fig:M67}. At a fixed metallicity of $Z=0.020$, the median age from fitting varies from log(age)=9.6 to log(age)=9.8, maintaining the trend that $\alpha$-enhanced compositions prefer older ages to match the observational data. When fixing the iron abundance, a slightly younger age is preferred when using a lower total $Z$, with most models preferring a log(age) of 9.7 (5\,Gyr), broadly consistent with literature age estimates.

Overall, across all three clusters, the best-fitting age when all compositions are considered are within 0.1\,dex of the existing best estimates in the literature. A consistent trend is found whereby, at fixed $Z$, an increase in \afe\ requires an older stellar population in order to generate an isochrone which matches the observations, with a typical uncertainty of $\pm0.1-0.2$\,dex. This highlights the level of uncertainty introduced in cluster age determination if robust metallicity (and crucially \afe) measurements are lacking. The shift towards an older population to maintain the same Main Sequence turn-off suggests that the change in $T_\mathrm{eff}$ from an increase in \afe\ has a stronger impact on the isochrone than the change in Main Sequence lifetime. If the change of lifetime was the dominant factor, the shorter evolutionary timescales would require a younger stellar population to have the same turn-off point.

\subsection{Comparison to BaSTI theoretical isochrones}
\label{sec:basti}
\begin{figure}
    \centering
    \includegraphics[width=0.48\textwidth]{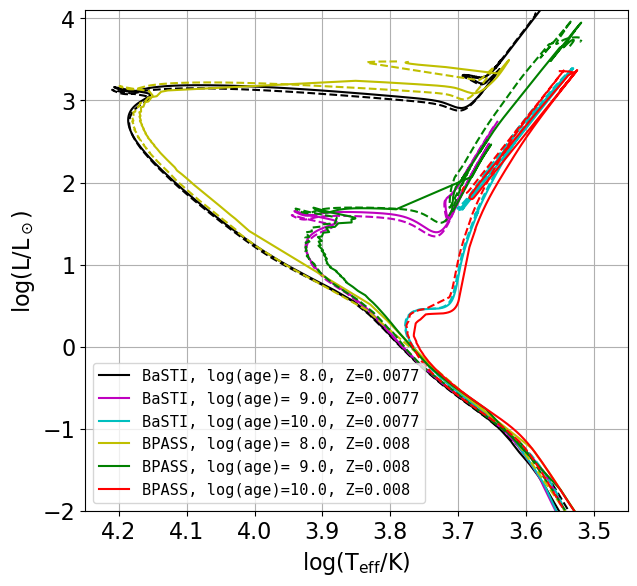}
    \caption{Comparison of solar-scaled (solid lines) and $\alpha$-enhanced isochrones (dashed lines) at 100 Myr, 1 Gyr and 10 Gyr, from BaSTI and from the \bpass\ models in this work.}
    \label{fig:BaSTIBPASS}
\end{figure}

While the literature lacks large grids of $\alpha$-enhanced stellar evolution models at high mass, there are a number of works at low stellar mass, used in the context of globular clusters. Old metal-poor populations such as globular clusters represent the low-mass survivors of the $\alpha$-enhanced stellar populations that formed in the early Universe. Here we compare our results with the existing BaSTI stellar evolution model grid, which explored the effect of $\alpha$-enhancement in recent data releases \citep{Pietrinferni21}. From BaSTI, we select Solar-scaled isochrones with \feh=-0.30 ($Z=0.00770$) and isochrones with \afe=+0.4 and \feh=-0.6 ($Z=0.00773$). We compare these to the \bpass\ models at $Z=0.008$ in Fig~\ref{fig:BaSTIBPASS}. The solid lines indicate Solar-scaled isochrones at 100 Myr, 1 Gyr and 10 Gyr, while the equivalent $\alpha$-enhanced isochrones are shown as dashed lines. At all three ages, the BaSTI isochrones look nearly identical along the Main Sequence, with only the transition from the Main Sequence to the red giant branch through the Hertzsprung gap showing noticeable deviations. Conversely, the \bpass\ isochrones consistently show an offset between the Solar composition curves and the $\alpha$-enhanced ones, with the $\alpha$-enhanced curves being generally shifted to hotter temperatures, as expected based on the shifts described for the individual stellar models in earlier sections.

Notably, the $\alpha$-enhanced \bpass\ isochrones overlap well with both sets of BaSTI isochrones. One probable source for this is the different assumed Solar compositions. BaSTI uses the composition of \citet{Caffau11} as their Solar reference, as opposed to the \citet{Grevesse93} abundances used in \bpass. The \citet{Grevesse93} abundances are slightly more oxygen-rich, with the oxygen-to-iron ratio being 0.13 dex higher. Thus, a mildly $\alpha$-enhanced \bpass\ model would be expected to have a similar O/Fe ratio to a Solar-scaled BaSTI model. Another point of note is that when \citet{Pietrinferni21} compared their isochrones to earlier work, the DSEP database \citep{Dotter08}, they showed that the DSEP models were systematically slightly cooler on the MS than the BaSTI models and the subgiant branch of the DSEP models were slightly less luminous.

This appears broadly consistent with the offsets we see between our \bpass\ models and BaSTI. The DSEP models use the \citet{Grevesse98} Solar abundances, which is a small refinement of the earlier \citet{Grevesse93} abundances used in \bpass. This suggests that much of the offset between BaSTI and \bpass\ isochrones can be attributed to the differences in assumed Solar composition. Other differences such as the use of bespoke OPAL opacity tables for each composition in \bpass\ may explain the larger offset seen between the $\alpha$-enhanced and Solar-scaled isochrones.

\section{Discussion}

\subsection{Old, low-mass stellar populations}

Previous work such as \citet{Salaris98} has demonstrated how $\alpha$-enhanced stellar population isochrones are necessary to reliably determine the age of metal-rich globular clusters such as 47 Tucanae, leading to a reduction in the age estimates of three clusters by 0.8 Gyr from 10 Gyr to 9.2 Gyr, $\sim 0.036$ dex in log(age). Our work shows that at a metallicity of $Z=0.020$, the Main Sequence lifetimes of $\sim1$ \msun\ stars can be up to to 0.3 dex shorter at \afe$=+0.6$ (Fig~\ref{fig:ATAMS}, panel d). These shorter lifetimes will lead to stars turning off the Main Sequence earlier. Nevertheless, our $\alpha$-enhanced isochrones shown alongside the open cluster data (e.g. lower two panels of Fig~\ref{fig:M67}) clearly indicate shifts in the Main Sequence turn-off such that $\alpha$-enhanced models give an older age for the same turn-off point. This is due to the offset in effective temperature due to $\alpha$-enhancement being stronger than the age effect in the \bpass\ models. The \citet{Salaris98} models are a precursor to the more recent BaSTI models, and thus the differences in modelling technique such as the use of bespoke opacity tables in \bpass\, as discussed in Section~\ref{sec:basti} above will contribute to the difference in behaviour. From the indicative fits in Figs~\ref{fig:Pleiades}~-~\ref{fig:M67} and the isochrone comparisons in Fig~\ref{fig:BaSTIBPASS}, it is evident that the \bpass\ $\alpha$-enhanced isochrones can be up to 0.2 dex older in log(age) and match the Main Sequence turn-off of an equivalent Solar-scaled isochrone. 

\subsection{Young stellar populations}

The impact of $\alpha$-enhancement in the evolution of massive stars has not been explored in great detail in the literature. The works of \citet{Grasha21} and \citet{Farrell22} mentioned earlier are the most recent studies to explore the impact that composition changes have on massive star evolution. Our work echoes the overall conclusions of these earlier works, highlighting that changes in composition (at fixed metallicity) impact the evolution of stars in a variety of ways across the entire stellar mass range.

\citet{Agrawal22} and \citet{Eldridge22Review} highlight that sizeable uncertainties remain in massive star evolution without even considering changes in composition. The choice of stellar evolution code \citep[and the consequent physics choices such as wind mass loss rates,][]{2022ARA&A..60..203V} lead to large variations in luminosity, radius and remnant masses, particularly above 40 \msun. Uncertainties in stellar evolution and the underlying physics can have a marked effect on the behaviour of a young stellar population. 

In our work we have found that massive stars (8$\le$ M/\msun$\le$100), which dominate the light of young stellar populations, differ in their TAMS temperatures by about 0.1--0.2 dex when considering $\alpha$-enhancement, with a higher \afe\ being associated with a higher effective temperature. These differences become less pronounced ($\sim0.05$ dex) at 10 per cent Solar metallicity, which is more representative of what one expects to find in the distant Universe. Nonetheless, these differences provide another source of uncertainty in the ionizing output and ultraviolet spectral predictions of stellar models and should be borne in mind when fitting models to observations to infer stellar population properties.

\subsection{Impact on binary population and spectral synthesis}
\label{sec:bins}
The models presented in this work only consider single stars which have evolved in isolation. To better explore the impact that $\alpha$-enhancement has on a full population and spectral synthesis, one would need to consider binary evolution models. The light from young stellar populations is dominated by massive stars, which typically have high companion fractions \citep{2012Sci...337..444S}, so this is particularly pertinent to young, high-redshift galaxies.

Despite the lack of binary models at present, the changes in the evolution of single stars as a result of $\alpha$-enhancement can provide an indication of the changes which might be expected.
\begin{enumerate}
    \item At low metallicity, $\alpha$-enhancement leads to smaller radii at the end of the Main Sequence (lower right panel of Fig~\ref{fig:RTAMS}).
    \item TAMS ages are largely unaffected by $\alpha$-enhancement for low-metallicity, high-mass stars (lower-right panel of Fig~\ref{fig:ATAMS}), while effective temperatures are higher (Fig~\ref{fig:TTAMS}).
    \item At low masses, $\alpha$-enhanced models show shorter TAMS ages. This will impact the delay times in Type Ia SNe, since the evolution will proceed more rapidly.
    \item Smaller radii suggest that binary interactions would be less common, since the stars are more compact and would less readily fill their Roche lobes. This may reduce the number of rejuvenated stars and make the stellar population as a whole slightly redder at ages where binary interactions dominate.
    \item The unchanged luminosities but higher effective temperatures indicate that $\alpha$-enhanced stars would be expected to have bluer spectra than their Solar-scaled counterparts.
\end{enumerate}  

The latter two points indicate competing effects, changing the colour of the integrated light in opposing directions. This emphasises the need for a full grid of $\alpha$-enhanced binary star calculations including evolution models and stellar atmosphere models in order to fully explore the effect that $\alpha$-enhancement has on stellar populations, particularly young populations in the distant Universe.  Computing a complete grid of binary stellar evolution models, covering a range of metallicities, \afe\ values, orbital separations and mass ratios is a time-consuming process. Work in this area is ongoing, both in terms of binary evolution modelling (Byrne et al., in prep.) and in stellar atmosphere calculations (Stanway et al., in prep.).

Binary stellar evolution calculations by \citet{Nguyen24} show that interacting massive binaries where the primary has developed a helium core are a key contributor to the enrichment of material in stellar clusters. This enrichment process is necessary to explain the presence of multiple abundance patterns in stellar populations in massive clusters. As the frequency of interactions, and the evolutionary timescales on which they take place are sensitive to \afe, future detailed binary grids will provide further indication of the contribution that interacting binaries make to the enrichment of the material in stellar clusters.

\section{Conclusions}

In this work we have presented a grid of single-star stellar evolution models computed within the \bpass\ framework, incorporating the effect of $\alpha$-enhancement in massive stars for the first time. While the ultimate goal is to produce binary evolution models for applications to stellar populations in the distant Universe, the single-star results yield a number of important results.

\begin{enumerate}
    \item The main sequence lifetime of metal-rich low-mass stars ($<$8\,\msun) shows a strong dependence on $\alpha$-enhancement when the bulk metallicity is held fixed, such that $\alpha$-enhanced stars have shorter lifetimes. At Solar metallicity, $\Delta$\afe=0.6 results in up to 0.3 dex difference in Main Sequence lifetime. At a tenth solar, this is reduced to less than 0.1 dex. The lifetimes of massive stars (8$<$M/\msun$<$100) are largely independent of $\alpha$-enhancement.  

    \item  Temperature and radius at the end of the Main Sequence is dependent on the stellar composition at all masses. These properties show a systematic trend with $\alpha$-enhancement, towards shorter radii and higher temperatures, although typically by $<$0.1\,dex across most of the mass range and the metallicities considered here.

    \item Trends in the TAMS physical properties of models at fixed total metal mass fraction but varying composition are dominated in all cases by the depletion of iron as $\alpha$ elements are enhanced. However, even when controlling for this, weaker trends in properties continue to be observed. 

    \item RSG SN progenitor luminosities show slight sensitivity to $\alpha$-enhancement, but the changes are smaller than the current observational uncertainties for SN progenitor stars. 
    
    \item At high metallicity mass fraction the final mass, final core mass and inferred remnant masses are strongly dependent on $\alpha$-enhancement. This will impact on predictions for transients arising from the stellar remnant population. The dependence weakens at lower metallicity.

    \item The minimum mass of a star expected to end its life as a CCSN as a function of [Fe/H] is largely independent of \afe. On the other hand, this minimum mass varies considerably for a fixed value of \oxh\ when considering \afe\ variation. This indicates that large uncertainties could occur when estimating the number and rate of SNe expected if \oxh\ is used as a proxy for metallicity.

    \item Comparing isochrones to the stellar properties observed in nearby open clusters, the models match well to the observations. Changes in \afe\ can lead to a differences in inferred cluster ages of $\pm0.1-0.2$\,dex in log(age/yr).

    \item In the very massive star (M$>100$\,\msun) regime the evolution in all properties is largely dominated by extreme mass loss rates due to radiatively-driven winds. These are strongly metallicity dependent, and largely driven by iron. Future work will further explore their dependence on stellar composition.
    
    \item A complete set of binary evolution models is needed to draw firmer conclusions on young stellar populations, particularly for very massive stars and population ages in which binary interactions play a significant role. This is an ongoing project, but will be informed by the trends seen in the single star population.

\end{enumerate}

\section*{Acknowledgements}

We thank the anonymous reviewer for their feedback which has led to improvements throughout the manuscript. We thank members of the \bpass\ team for providing helpful discussions and feedback while carrying out this research. The {\sc{hoki}} package \citep{HOKI} was used to extract and manipulate output data from the \bpass\ stellar evolution models.
Stellar parameter data from SDSS and LAMOST catalogues was extracted using TOPCAT \citep[][]{TOPCAT1,TOPCAT2}. ERS and CMB acknowledge funding from the UK Science and Technology Facilities Council (STFC) through Consolidated Grant Numbers ST/T000406/1, ST/X001121/1. JJE acknowledges funding from the Royal Society Te Apar\=angi of New Zealand Marsden Grant Scheme. 
Computing facilities were provided by the Scientific Computing Research Technology Platform (SCRTP) of the University of Warwick. Data processing made use of Astropy\footnote{\url{https://www.astropy.org/}}, a community-developed core Python package for Astronomy \citep{astropy:2013,astropy:2018}. 

\section*{Data Availability}

Abridged versions of these \bpass\ v2.4.0 stellar model outputs will be made available at the \bpass\ websites\footnote{\url{https://warwick.ac.uk/bpass}}\footnote{\url{https://bpass.auckland.ac.nz}}. Detailed outputs will be made available upon reasonable request to the authors.



\bibliographystyle{mnras}
\bibliography{mybib} 




\bsp	
\label{lastpage}

\clearpage


\appendix
\section*{}

\noindent {\Large{\textbf{BPASS stellar evolution models incorporating $\alpha$-enhanced composition -- \\ I. Single star models from $0.1$ to $316$\,\msun}}}
\vspace*{11pt}
\section{Supplementary Information}
\pagenumbering{roman}
\subsection{Radius changes with change in composition}
\label{app:RTAMS}

\begin{figure*}
    \centering
    \includegraphics[width=\textwidth]{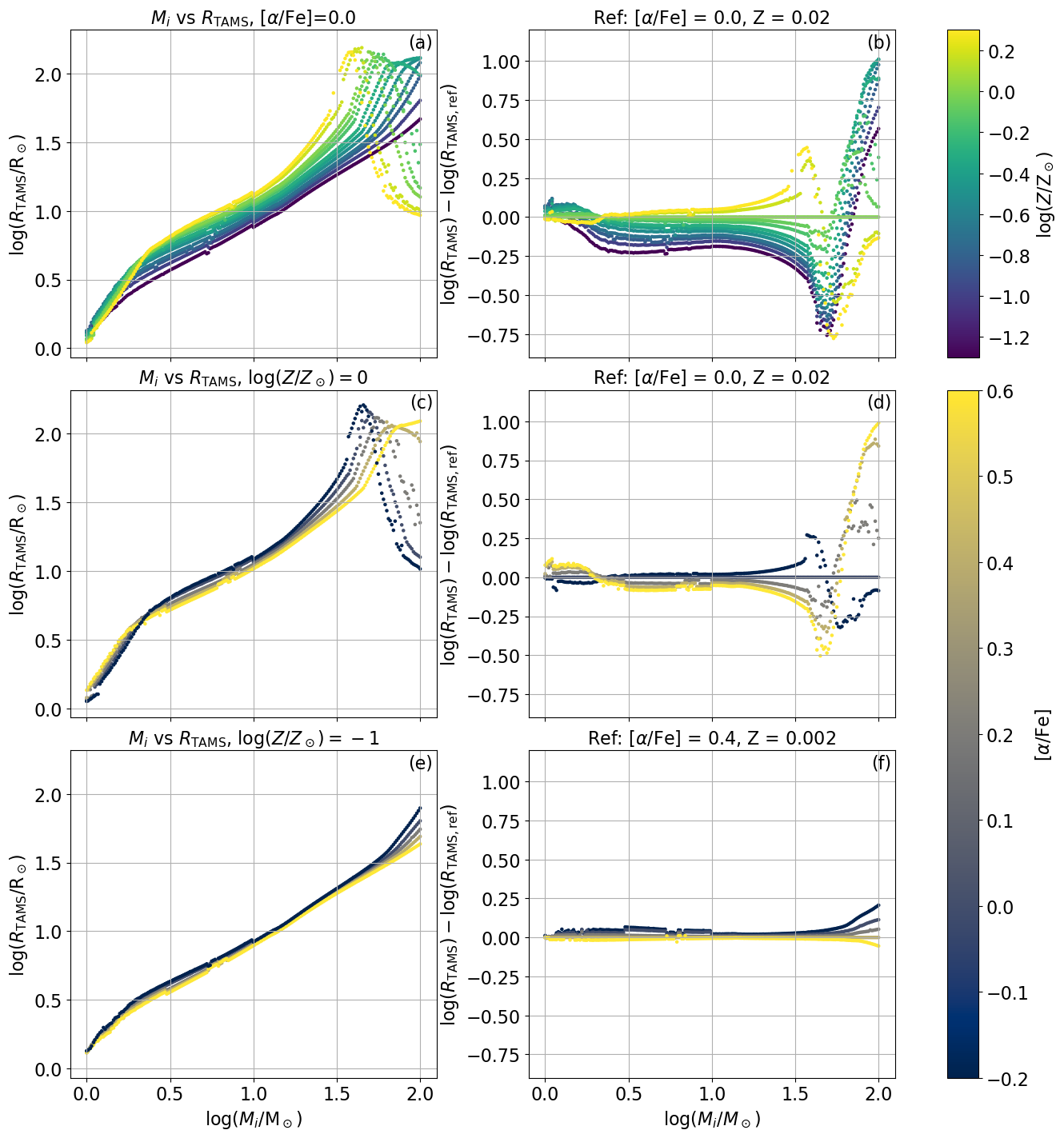}
    \caption{As Fig~\protect\ref{fig:ATAMS}, but for TAMS radii, rather than age.}
    \label{fig:RTAMS}
\end{figure*}

Fig~\ref{fig:RTAMS} is a similar plot to Fig~\ref{fig:ATAMS}, but for stellar radii at the end of the Main Sequence. The upper panels, (a)
and (b), show a large dispersion resulting from changes in $Z$, with metal-poor stars having smaller radii at the end of the Main Sequence. A similar trend is seen for fixed $Z$ and variable \afe, seen in panels (c) and (d). Like the other stellar properties, the changes are qualitatively similar between the $\alpha$-enriched, iron-poor compositions and low bulk-metallicity cases, indicating that iron and iron opacity are playing a large role in determining the stellar radius. The $\alpha$-sensitive effects are minimal at low $Z$, as can be seen in panels (e) and (f). Radii changes may be crucial when one considers binary systems, since a star with a larger radius will more easily overfill its Roche lobe. 

\subsection{Comparative diagrams for compositions matched on iron abundances}
\label{app:full}
\label{app:samefe}

\begin{figure*}
    \centering
    \includegraphics[width=0.97\textwidth]{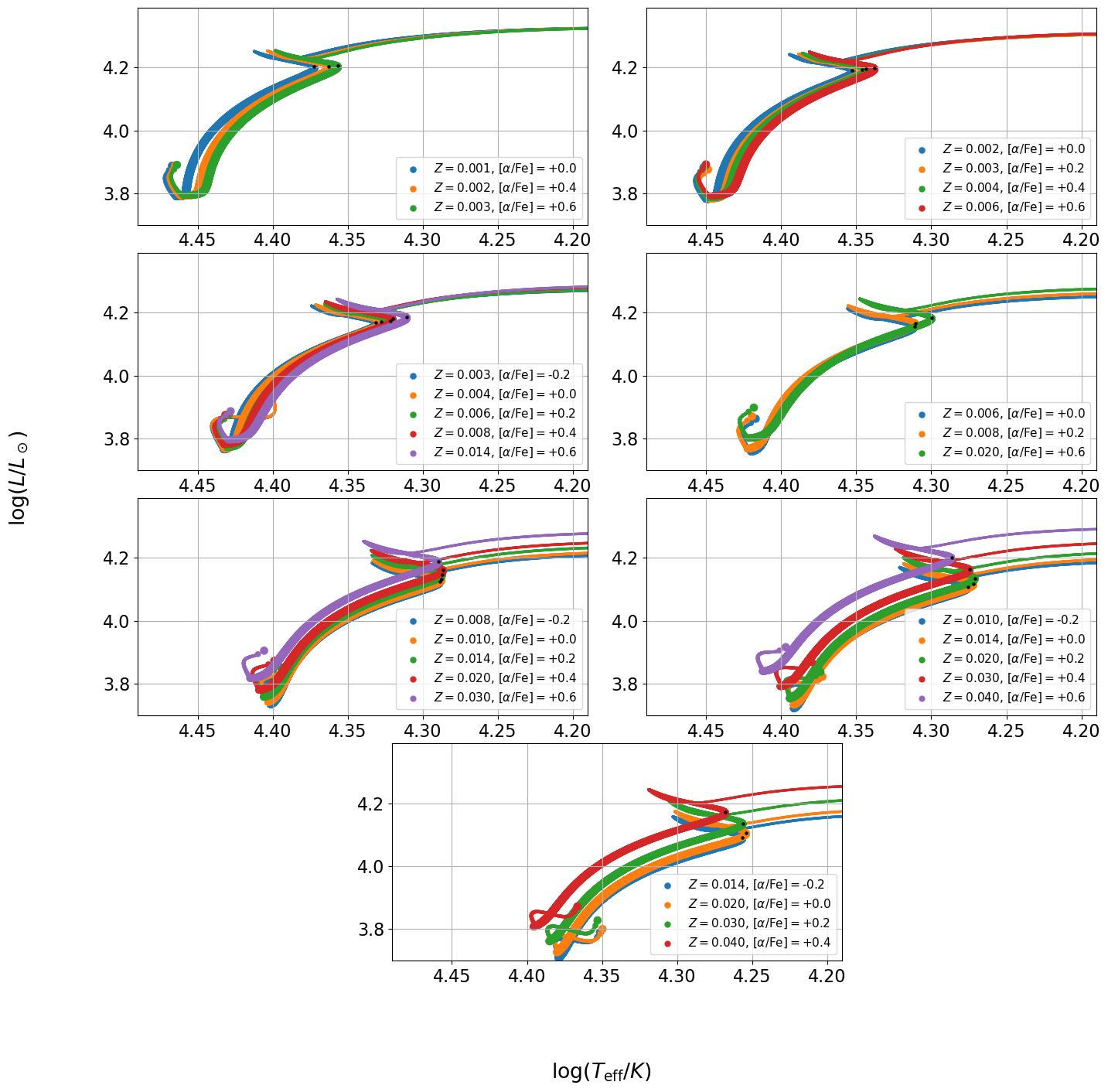}
    \caption{HRDs  showing the Main Sequence evolution of 20 \msun models. Each panel shows a selection of compositions chosen to have near-identical iron mass fractions.}
    \label{fig:sameFe_HR}
\end{figure*}

Fig~\ref{fig:sameFe_HR} shows comparative HRDs for 20 \msun\ models with matching iron abundances. Each panel shows a selection of compositions which have a specific iron abundance, as listed in Table~\ref{tab:sameFe}. As discussed in Section~\ref{sec:fixFe}, given the discrete nature of the composition grid, these do not match perfectly, but are within a few per cent of the same iron abundance. Small differences remain in the evolution tracks shown here. This indicates that even when one fixes the iron abundance, changes in \afe\ (and the resulting stellar opacity tables) still contribute to minor shifts in the stellar evolution tracks. 

\begin{figure*}
    \centering
    \includegraphics[width=0.95\textwidth]{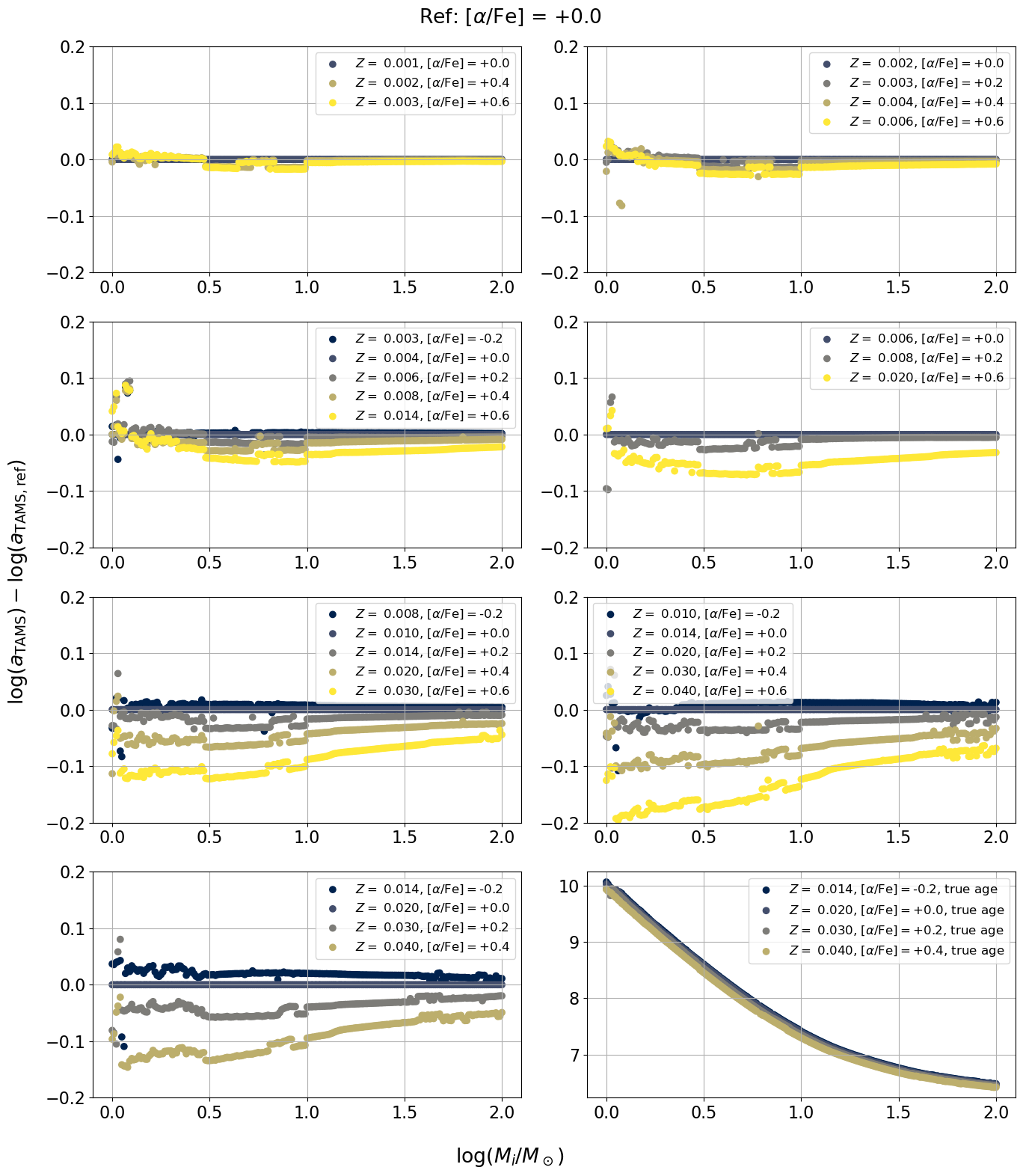}
    \caption{Differences (in log space) in TAMS age between compositions with near-identical iron mass fractions, relative to the \afe=0 composition in each panel. The final (lower right) panel compares the actual ages of the final set, rather than the relative differences.}
    \label{fig:sameFe_A}
\end{figure*}

\begin{figure*}
    \centering
    \includegraphics[width=0.95\textwidth]{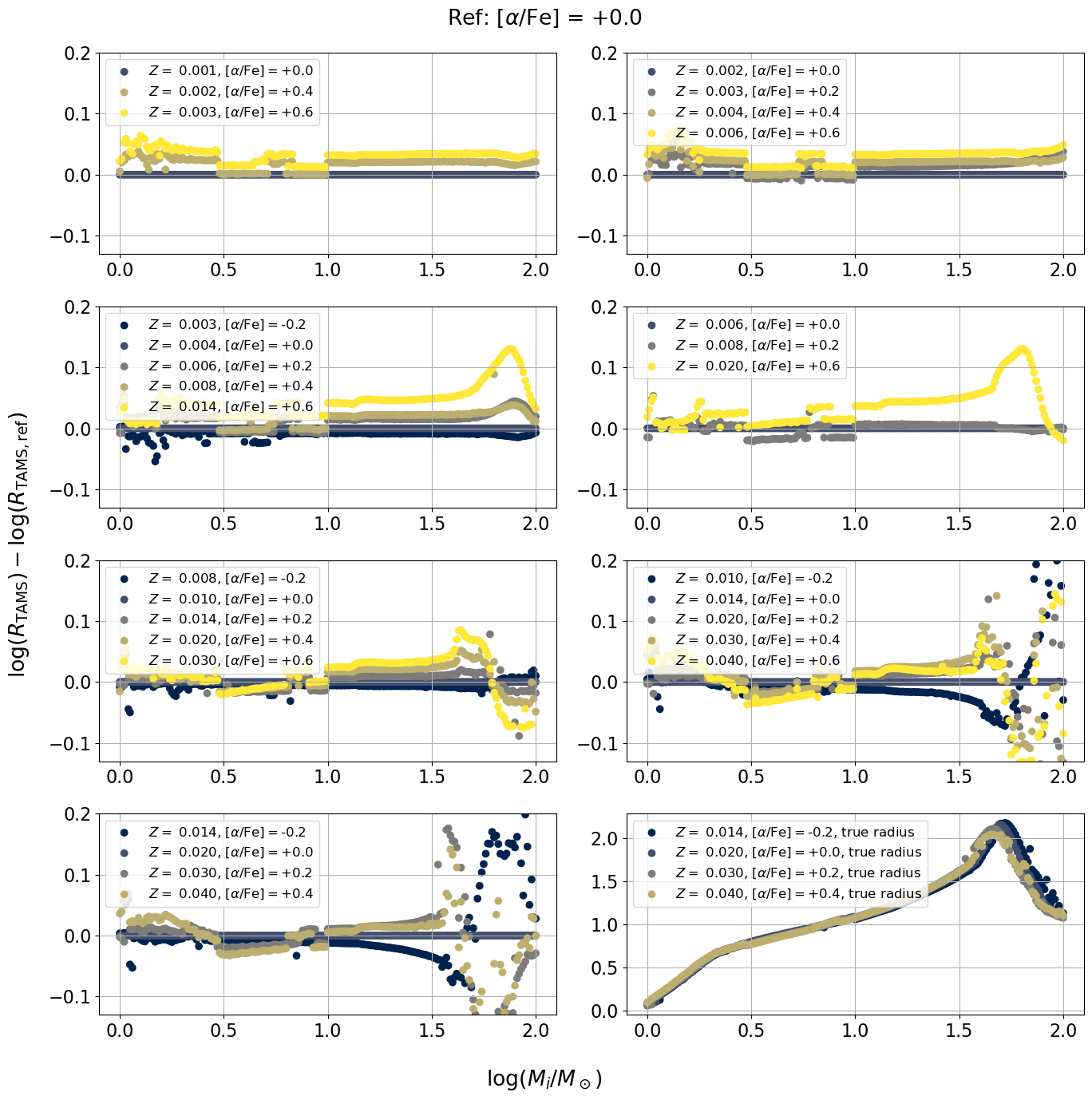}
    \caption{As Fig~\protect\ref{fig:sameFe_A}, but for $\log(R_{\mathrm{TAMS}})$}
    \label{fig:sameFe_R}
\end{figure*}

Figs~\ref{fig:sameFe_A} and~\ref{fig:sameFe_R} show the relative differences in log(age/yr) and log(R$_{\mathrm{TAMS}}$/\rsun) respectively. Once again, there are some small differences seen for each set of iron abundances, with $\alpha$-enhanced models showing a slight preference for larger radii and shorter Main Sequence lifetimes.


\subsection{Massive star end states}
\label{app:SNe}

\begin{figure*}
    \centering
    \includegraphics[width=0.97\textwidth]{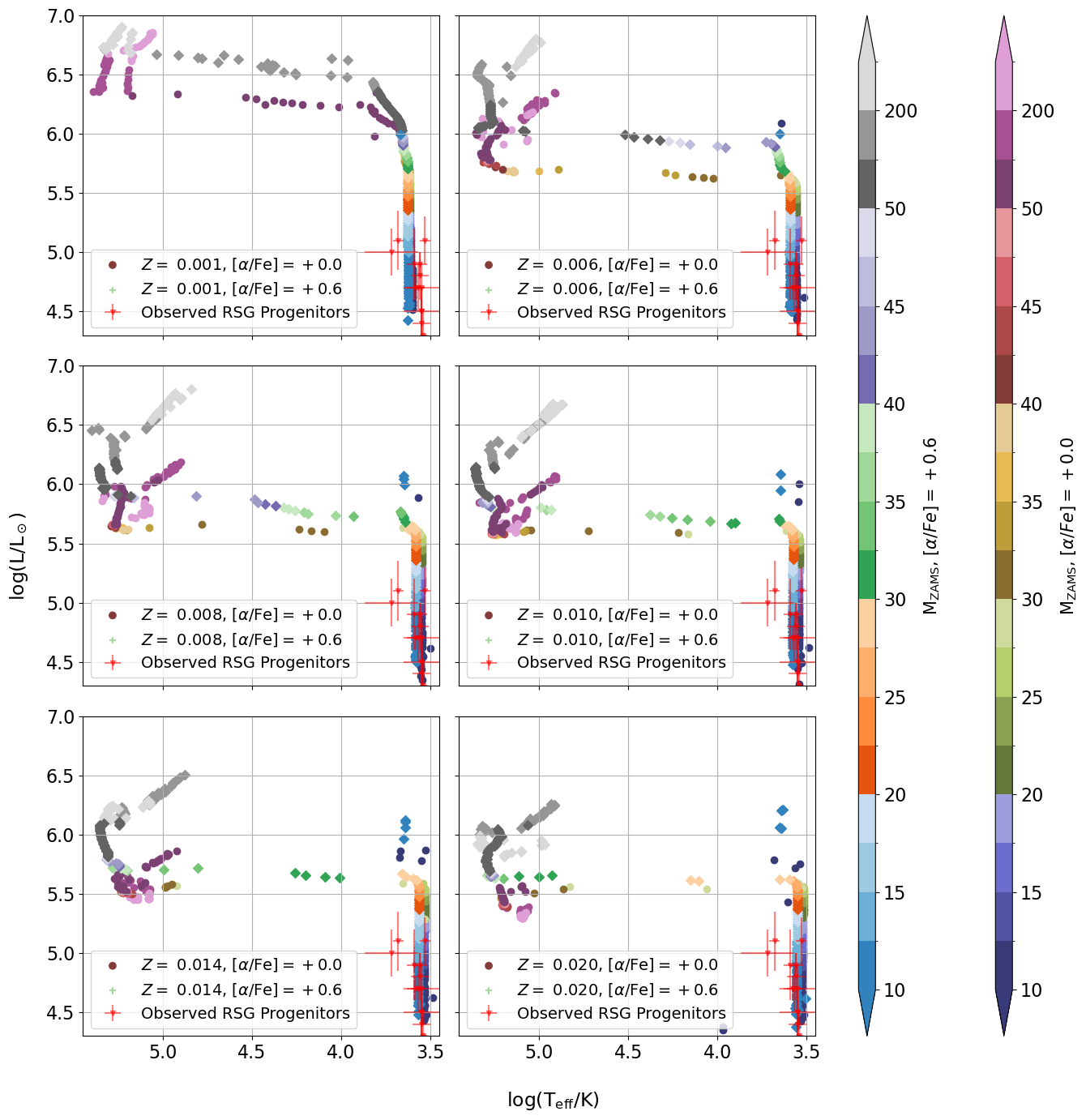}
    \caption{HRDs showing the locus of massive stellar models (M$_{\mathrm{ZAMS}}$/M$_\odot$>8) at the end of their evolution, compared to observed SN progenitors from \protect\citet{Smartt15} and \protect\citet{ONeill21} (red triangles). Each panel corresponds to a different total $Z$, from $Z=0.001$ in the upper left panel to $Z=0.020$ in the lower right panel. The models with \afe$=0.0$ are represented by the circles, coloured by initial mass. The coloured diamonds correspond to the models with \afe$=+0.6$. Each step in the respective colour scales represents an increment of 2.5 \msun, up to 50 \msun, with large increments of 75 \msun\ above this.}
    \label{fig:SNe_HRD}
\end{figure*}

Fig~\ref{fig:SNe_HRD} shows the final location on the HRD of stars with masses greater than 8 \msun, both at Solar-scaled composition and for an $\alpha$-enhancement of $+0.6$. Each panel corresponds to a different bulk metallicity, $Z$, ranging from $Z=0.001$ in the upper left panel to $Z=0.020$ in the lower right panel. On the right hand-side of each panel we can see the low-temperature RSGs, the stars with the lowest initial masses in this mass range.

Generally speaking, there are only minor differences between the Solar-scaled and $\alpha$-enhanced models. At intermediate metallicities ($Z=0.006$, upper-right panel and $Z=0.008$, centre-left panel) the $\alpha$-enhanced RSGs extend to noticeably higher luminosities and masses. At higher metallicity, mass-loss rates are high, and the critical mass at which stars are stripped and become BSGs is reduced in both cases. At low metallicity, mass-loss rates are much lower, and the relative difference in composition is very small, so there is little change in their evolution.

As stars which are the product of stripping through mass-loss and/or binary interactions, the behaviour of the hot, massive, BSGs requires binary evolution models and a clear understanding of the appropriate mass-loss schemes to draw definitive conclusions on the overall impact of $\alpha$-enhancement.

\begin{figure}
    \centering
    \includegraphics[width=0.48\textwidth]{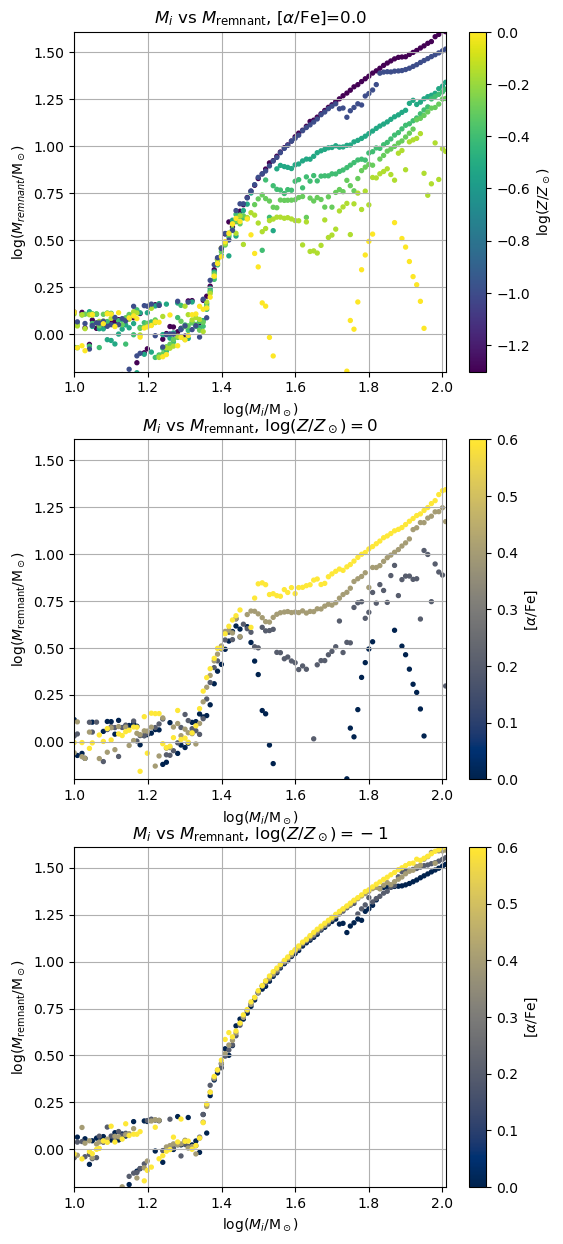}
    \caption{As Fig~\protect\ref{fig:MFINA}, but for the stellar remnant masses predicted using the \bpass\ approach.}
    \label{fig:SNe_M_rem}
\end{figure}

Fig~\ref{fig:SNe_M_rem} uses the same combination of panels and compositions as Fig~\ref{fig:MFINA}, but considering the remnant mass at the end of the stellar evolution calculation. The remnant mass is calculated based on the amount of mass remaining after $10^{44}$\,J of energy has been ejected into the stellar envelope as per \citet{Eldridge04b}. These panels show broad agreement with the results for CO core masses shown in the right-hand panels of Fig~\ref{fig:MFINA}. Models with low $Z$ and/or high \afe\ show larger remnant masses, although the behaviour is less smooth and shows more extreme variations than seen for the CO core masses.

\subsection{Fitting to stellar clusters}
\label{app:clusfit}

\begin{figure}
    \centering
    \includegraphics[width=0.48\textwidth]{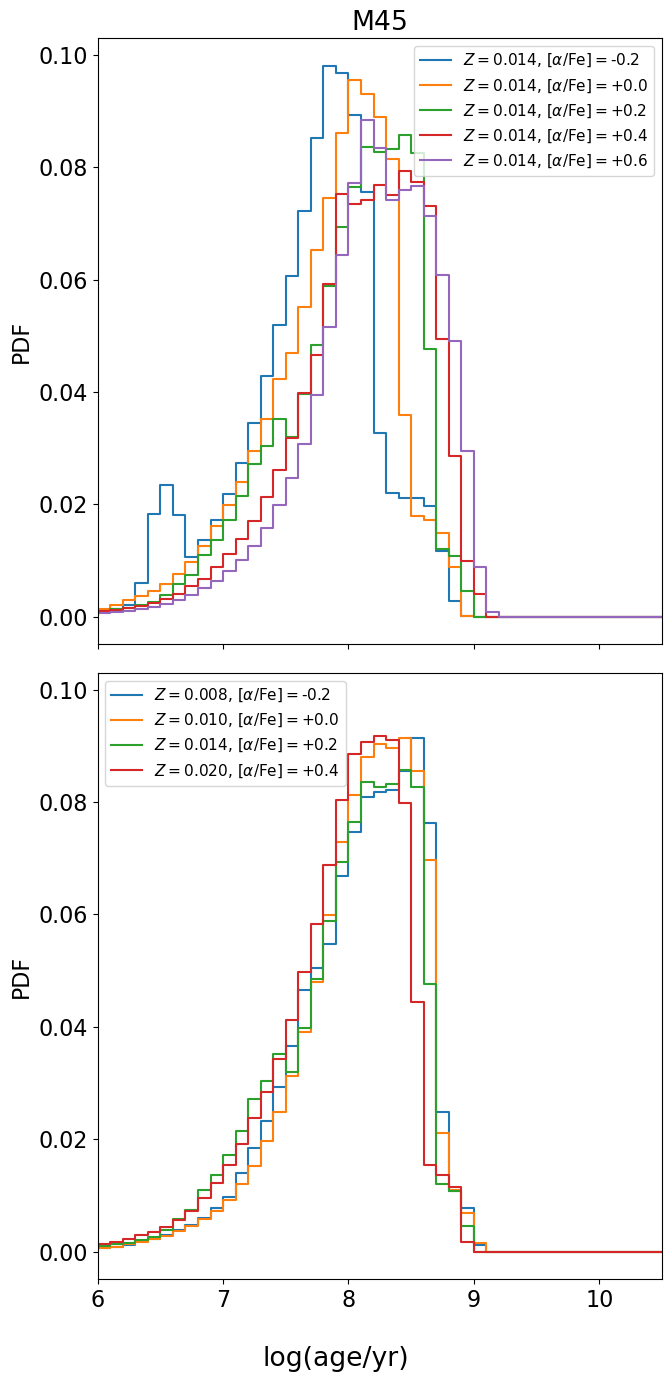}
    \caption{Probability density functions showing the best fitting stellar population ages from Age Wizard for selected metallicities and $\alpha$-enhancement factors. Upper panel: A fixed total metallicity mass fraction of $Z=0.014$, with variation in \afe. Lower panel: Varying $Z$ and \afe\ to maintain a near constant Fe abundance of about 0.5 times the Solar value.}
    \label{fig:AWiz_M45}
\end{figure}

\begin{figure}
    \centering
    \includegraphics[width=0.48\textwidth]{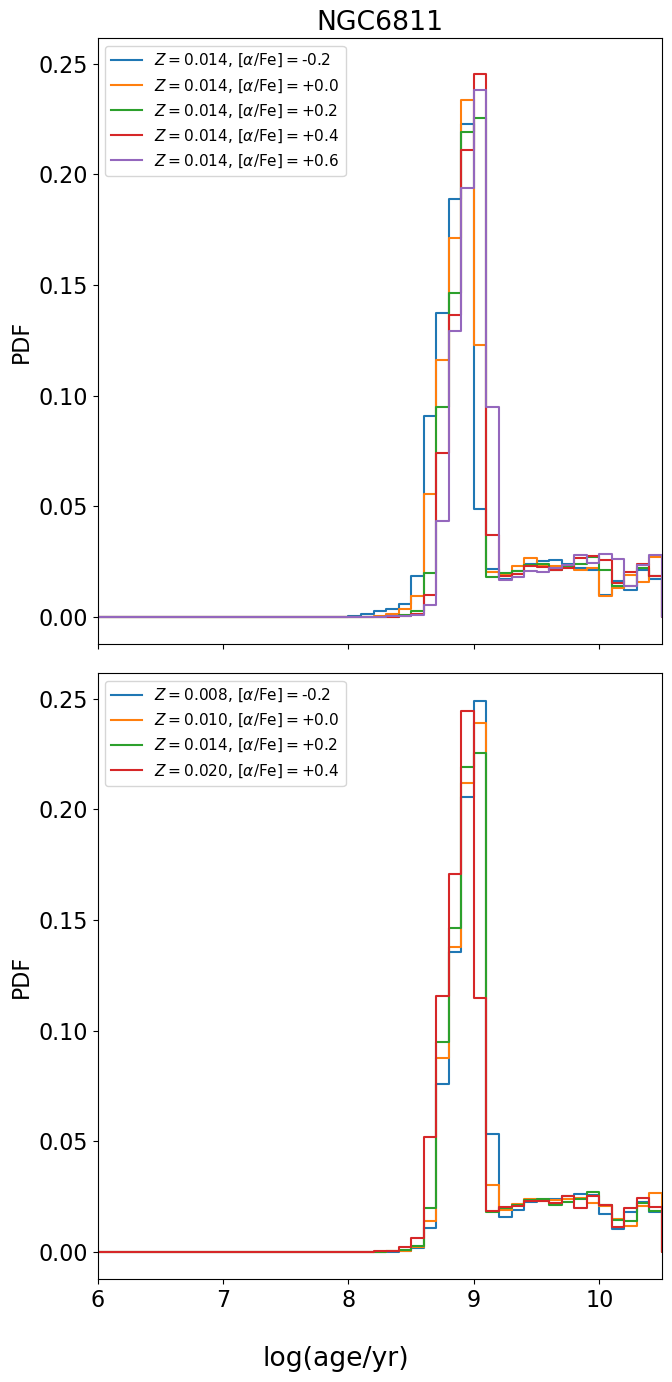}
    \caption{As Fig~\protect\ref{fig:AWiz_M45}, but for NGC~6811.}
    \label{fig:AWiz_NGC6811}
\end{figure}

\begin{figure}
    \centering
    \includegraphics[width=0.48\textwidth]{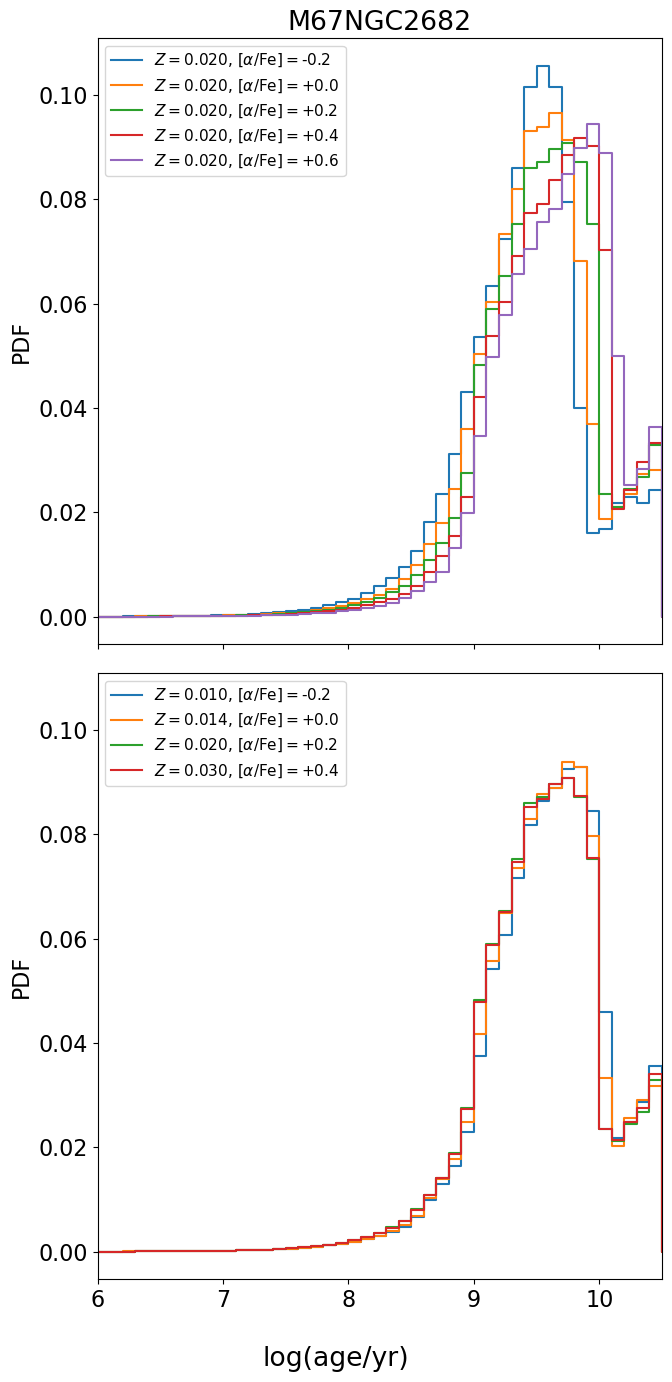}
    \caption{As Fig~\protect\ref{fig:AWiz_M45}, but for NGC~2682. The upper panel considers a metallicity mass fraction of $Z=0.020$ in this case.}
    \label{fig:AWiz_NGC2682}
\end{figure}

Figs~\ref{fig:AWiz_M45}, \ref{fig:AWiz_NGC6811} and~\ref{fig:AWiz_NGC2682} show the results of using the Age Wizard tool in {\sc{hoki}} \citep[][]{HOKI,Stevance20AgeWizard} to determine the log(g)--log(T) isochrones which best fit the observational data for M45/Pleiades/Matariki, NGC~6811 and M67/NGC~2682 respectively.

For M45 (Fig~\ref{fig:AWiz_M45}), in order to get a reliable fit, we had to limit the observational data to the hottest stars ($\log(T_\mathrm{eff})/\mathrm{K} > 3.8$. Including all of the cooler stars leads to spurious age fits of $10^{10.5}$\,yr for the cluster. This is due to the number of points at low effective temperature and the fact that Age Wizard is designed to take the density of points into account, as is necessary in a binary stellar population where not all stars lie on a neat line at a fixed age. Effectively, the fitting considers the comparatively small number of high temperature stars as outliers and proceeds to fit the dense cluster of low-temperature stars as an extremely old stellar population. Constraining the sample to the high temperature stars forces the Age Wizard to fit them appropriately, considering them as Main Sequence stars. Doing so yields a probability density function which is relatively broad, but with a peak close to 100\,Myr. The broad peak highlights the lack of constraining power when there are no post-Main Sequence objects in the sample. There is a slight preference for older ages when $\alpha$-enhanced models are considered, with the peak of the probability density varying from $10^{7.9}$\,yr to $10^{8.5}$\,yr between the different compositions at $Z=0.014$. Choosing combinations with matched iron abundance (lower panel) the variation in the probability density functions is much lower, again indicating that iron is the predominant element affecting the age determination of a stellar population.

For NGC6811 (Fig~\ref{fig:AWiz_NGC6811}), the best fitting age shows a strong peak at $10^9$ or $10^{9.1}$\,yr, showing very weak dependence on \afe\ or \feh. A number of factors combine to give this result. The change in stellar lifetimes for a fixed mass ($\alpha$-enhanced stars with shorter lives, thus a higher implied cluster age for a given MS turn-off point) is offset by the overall shift in effective temperature ($\alpha$-enhanced stars are hotter, thus a younger implied age for a given Main Sequence turn-off). The presence of some post-main sequence stars in the cluster also provide an additional extra constraint, putting a firmer lower limit on the age.

In M67/NGC2682 (Fig~\ref{fig:AWiz_NGC2682}) varies from $10^{9.6}$\,yr at \afe$=-0.2$ to $10^{10}$\,yr at \afe$=+0.6$, once again showing a tendency for an older inferred cluster age if $\alpha$-enhancment is assumed. The change in probability density is much smaller when Fe abundance is kept fixed, with the peak only changing between $10^{9.9}$\,yr and $10^{10}$\,yr between the abundances plotted.


\end{document}